\documentclass[universe,article,accept,pdftex,moreauthors]{Definitions/mdpi} 
\makeatletter
\renewcommand{\maketag@@@}[1]{\hbox{\m@th\normalsize\normalfont#1}}
\makeatother

\firstpage{1} 
\pubvolume{1}
\issuenum{1}
\articlenumber{0}
\pubyear{2026}
\copyrightyear{2026}
\externaleditor{~} 
\datereceived{29 July 2026} 
\daterevised{28 August 2026} 
\dateaccepted{30 August 2026} 
\datepublished{ } 

\Title{Gravitational Baryogenesis in Energy-Momentum Squared~Gravity}

\newcommand\orcidDavid{{\href{https://orcid.org/0009-0006-8485-3140}{\orcidicon}}}

\newcommand{\orcidJPMimoso}{{\href{https://orcid.org/0000-0002-9758-3366}{\orcidicon}}} 
\newcommand\orcidFrancisco{{\href{https://orcid.org/0000-0002-9388-8373}{\orcidicon}}}

\Author{{David S. Pereira}
 $^{1,2,}$*\orcidDavid, Francisco S. N. Lobo $^{1,2}$\orcidFrancisco, José Pedro Mimoso $^{1,2,}$*\orcidJPMimoso}

\AuthorNames{David S. Pereira, Francisco S. N. Lobo, José Pedro Mimoso}

\address{%
$^{1}$ \quad Instituto de Astrof\'{\i}sica e Ci\^{e}ncias do Espa\c{c}o, Faculdade de
Ci\^encias {da} Universidade de Lisboa, Campo~Grande, Edif\'{\i}cio C8, 
P-1749-016 Lisbon, Portugal; {fslobo@ciencias.ulisboa.pt}\\
$^{2}$ \quad Departamento de F\'{i}sica, Faculdade de Ci\^{e}ncias {da} Universidade de Lisboa, Campo Grande, Edif\'{\i}cio C8, P-1749-016 Lisbon, Portugal}

\corres{Correspondence: {djpereira@ciencias.ulisboa.pt}~(D.S.P.);  
 jpmimoso@ciencias.ulisboa.pt~(J.P.M.)}

\abstract{We demonstrate that the matter sector itself can drive baryogenesis in energy-momentum squared gravity, $f(R,\mathcal T^2)$ with $\mathcal T^2\equiv T_{\mu\nu}T^{\mu\nu}$. High-density matter corrections provide new time-dependent sources for the baryon asymmetry through derivative couplings to $\mathcal T^2$ and to the full combination $f(R,\mathcal T^2)$. Notably, the decoupling temperature is not treated as a free parameter; instead, it is fixed by the freeze-out of the Weinberg $B-L$-violating operator, directly linking the asymmetry to the light-neutrino mass scale and to the modified expansion history. Analyzing representative powers $n=1/4$, $n=1/2$, $n=5/8$, and $n=1$, we find that entropy evolution sharply distinguishes the models. The $n=1/4$ branch cannot serve as a self-contained radiation-era model, the $n=1/2$ branch is entropy-safe but too weak to reproduce the observed asymmetry, and the $n=1$ branch survives only after entropy dilution. Remarkably, the $n=5/8$ branch provides the cleanest viable realization, generating the observed baryon-to-entropy ratio with controlled effective-field-theory hierarchies.}


\begin{document}

\section{Introduction}\label{sec:intro}

One of the central open problems in modern cosmology is the origin of the observed matter--antimatter asymmetry of the Universe~\cite{wilczek1980cosmic}. This asymmetry is usually quantified by the baryon-to-entropy ratio
\begin{equation} Y_B\equiv \frac{n_b}{s} = \frac{n_B-n_{\bar B}}{s}, 
\end{equation}
where $n_b$ is the net baryon number density and $s$ is the entropy density of the radiation bath. Measurements from Cosmic Microwave Background anisotropies and Big Bang Nucleosynthesis constrain this quantity to be approximately~\cite{WMAP:2003ogi,Planck:2018vyg,Fields:2019pfx,ParticleDataGroup:2020ssz}
\begin{equation}\label{eq:YB_obs_intro} Y_B^{\rm obs} = (8.8\pm0.6)\times10^{-11}. 
\end{equation}

Although this number is small, it is crucial: it encodes the fact that the visible Universe contains an excess of matter over antimatter and, therefore, requires a dynamical explanation beyond a perfectly symmetric primordial~state.

The standard theoretical framework for generating such an asymmetry is based on the Sakharov conditions~\cite{Sakharov}, which require baryon-number violation, C and CP violation, and~departure from thermal equilibrium. In~the Standard Model, anomalous electroweak processes such as sphalerons violate $B+L$, providing an important ingredient for baryogenesis. However, conventional electroweak baryogenesis within the Standard Model faces well-known difficulties, including insufficient CP violation and the absence of a strongly first-order electroweak phase transition~\cite{Trodden:1998ym,Morrissey:2012db,Cline:2017jvp,Pereira:2023xiw}. These limitations motivate the investigation of baryogenesis mechanisms beyond the Standard Model and, in~particular, mechanisms in which gravitational or cosmological effects play a direct~role.

\begin{figure}[H]
	\includegraphics[width=0.5\textwidth]{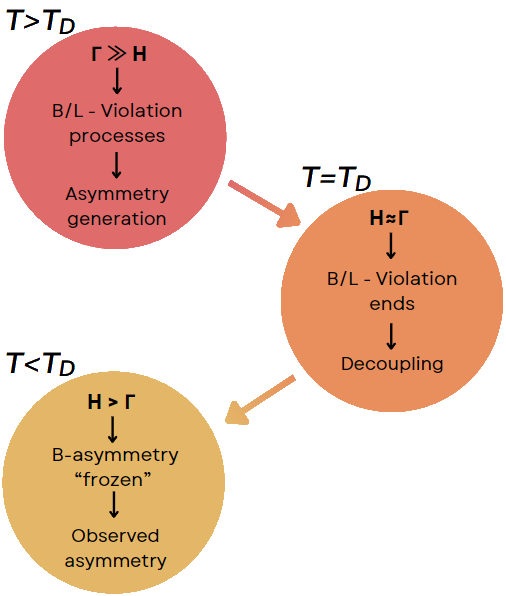}
	\caption{{Schematic}
	representation of baryon-asymmetry generation in thermal equilibrium. While the baryon- or lepton-number violating interaction rate satisfies $\Gamma\gg H$, the~asymmetry is generated in the thermal bath. When $\Gamma(T_D)\simeq H(T_D)$, the~relevant interaction decouples and the asymmetry freezes~out.}
	\label{Figure_1}
\end{figure}

The out-of-equilibrium condition can also be reconsidered. In~thermal equilibrium and under CPT invariance, particle and antiparticle distributions are identical, implying $n_B=n_{\bar B}$. However, if~CPT is dynamically or spontaneously violated in an expanding background, an~effective energy splitting between baryons and antibaryons may arise even in thermal equilibrium~\cite{Cohen:1987vi,Cohen:1988kt,Lambiase:2013haa,Dolgov:1991fr,Dolgov:1997qr,Arbuzova:2016cem}. In~this case, the~baryon asymmetry is generated while baryon- or lepton-number violating interactions remain efficient, and~it freezes when those interactions decouple. The~decoupling temperature $T_D$ is determined by
\begin{equation} \Gamma(T_D)\simeq H(T_D), \end{equation}
where $\Gamma$ is the interaction rate of the relevant baryon- or lepton-number violating process and $H$ is the Hubble expansion rate. This sequence is schematically illustrated in Fig.~\ref{Figure_1}. Therefore, $T_D$ is not an arbitrary input: it depends both on the microphysics responsible for $B$, $L$, or~$B-L$ violation and on the cosmological background determined by the gravitational theory.

Gravitational baryogenesis provides a particularly appealing realization of this idea. In~the original proposal of Davoudiasl~et~al.~\cite{Davoudiasl:2004gf}, the~interaction
\begin{equation} \frac{1}{M_\ast^2}\,\partial_\mu R\,J^\mu \end{equation}
couples the derivative of the Ricci scalar $R$ to a baryonic or $B-L$ current $J^\mu$. In~an expanding Universe, this term induces an effective chemical potential for particles and antiparticles and can generate a net asymmetry while the relevant number-violating interactions are still in thermal equilibrium. A~key feature of the mechanism is that the asymmetry is controlled by the time evolution of a gravitational scalar. Consequently, the~result is highly sensitive to the cosmological background and to possible high-energy deviations from General Relativity. This observation has motivated many extensions of gravitational baryogenesis in modified gravity frameworks~\cite{Lambiase:2006dq,Sadjadi:2007dx,MohseniSadjadi:2007qk,Lambiase:2012tn,Odintsov:2016hgc,Ramos:2017cot,Aghamohammadi:2017emk,Atazadeh:2018xjo,Bhattacharjee:2020wbh,Cruz:2025fuk,Pereira:2024ddu}. In~such scenarios, the~cosmological evolution may be modified, changing $R$, $\dot R$, and~$H(T)$, or~new scalar quantities arising from the gravitational sector may be coupled directly to $J^\mu$. This broader perspective suggests that baryogenesis can be used not only as a mechanism for explaining the matter--antimatter asymmetry, but~also as a probe of high-density gravitational~physics.

In this work, we investigate modified gravitational baryogenesis in the context of energy-momentum squared gravity (EMSG), also known as $f(R,T_{\mu\nu}T^{\mu\nu})$ gravity~\cite{Katirci:2013okf,Roshan:2016mbt}; see Ref.~\cite{Cipriano:2024jng} for a recent review. In~this theory, the~gravitational action depends on the Ricci scalar $R$ and on the self-contraction of the energy-momentum tensor,
\begin{equation} \mathcal T^2\equiv T_{\mu\nu}T^{\mu\nu}. \end{equation}

{The} scalar $\mathcal T^2$ becomes especially important at high energy density, making EMSG a natural framework for early-Universe cosmology. Since $\mathcal T^2$ scales quadratically with the matter sector, it can modify the Friedmann equations, the~time-temperature relation, and~the freeze-out condition for baryon- or lepton-number violating processes. These features make EMSG a well-suited arena for testing whether matter-sector modifications of gravity can enhance or suppress gravitational baryogenesis. A~distinctive aspect of EMSG and related matter-geometry theories is the presence of nonminimal couplings between matter and geometry. Similar structures appear in $f(R,\mathcal L_m)$ gravity~\cite{Bertolami:2007gv,Harko:2010mv}, $f(R,T)$ gravity~\cite{Harko:2011kv,Barrientos:2018cnx}, and~$f(R,T,R_{\mu\nu}T^{\mu\nu})$ gravity~\cite{Haghani:2013oma,Odintsov:2013iba}. In~EMSG, these effects are encoded in the additional scalar $\mathcal T^2$. While the theory reduces to an $f(R)$-type model in the absence of matter, deviations from General Relativity become relevant when the energy-momentum tensor is nonzero, especially in the high-density regime relevant for baryogenesis.

The purpose of this paper is twofold. First, we formulate modified gravitational baryogenesis in EMSG by considering derivative interactions involving the matter-sector scalar and the full gravitational function, schematically $\partial_\mu f(\mathcal T^2)J^\mu$ and $\partial_\mu f(R,\mathcal T^2)J^\mu$. Second, we determine whether the resulting asymmetry can reproduce Equation~\eqref{eq:YB_obs_intro} while satisfying the radiation-era dynamics, BBN constraints, and~effective-field-theory requirements of each model. A~further point of emphasis is the treatment of the decoupling temperature. Rather than taking $T_D$ as a free phenomenological parameter, we determine it from a concrete $B-L$-violating process. In~particular, we use the Weinberg operator, which induces $\Delta L=2$ interactions and relates the interaction rate to the light-neutrino mass scale. This connects the baryogenesis calculation directly to particle-physics data and makes the viability of each model more transparent.

Because EMSG contains nonminimal matter-geometry couplings, the~matter energy-momentum tensor is not generally conserved. Therefore, the~usual adiabatic interpretation of $n_b/s$ requires additional care. We distinguish the local thermal-equilibrium relation \mbox{$\rho=(\pi^2/30)g_\ast T^4$} from the stronger assumption of adiabatic radiation evolution. This leads us to track the modified continuity factor $C_{\rm con}$ and the possible entropy correction $\Delta_S$ relating the freeze-out asymmetry to the final asymmetry. This entropy analysis is essential for deciding which branches of the theory are self-consistent as early-Universe radiation-era models. { We focus on the energy-momentum powered subclass of EMSG, in~which the modification to the gravitational action has a power-law dependence on the energy-momentum-squared scalar, proportional to $(\mathcal{T}^2)^n$. The~exponent $n$, therefore, characterizes the particular power-law model under consideration and controls the dependence of the modified-gravity contribution on the matter energy density. We consider the representative~values
\[
n=\frac14,\qquad n=\frac12,\qquad n=\frac58,\qquad n=1.
\]}

{These} values lead to qualitatively different cosmological and thermodynamic behaviour. These values lead to qualitatively different cosmological and thermodynamic behaviour. The~case $n=1/4$ can generate a sizeable freeze-out asymmetry but has a non-adiabatic correction that grows as the Universe cools. The~case $n=1/2$ remains close to adiabatic radiation evolution but does not generate a sufficient asymmetry in the BBN-allowed region. The~value $n=5/8$ is singled out by the structure of the modified continuity equation~\cite{Board:2017ign}; for radiation, it gives the standard continuity law and, therefore, provides an entropy-safe branch. Finally, the~$n=1$ model can generate the observed asymmetry after including the calculable entropy-dilution factor.

This paper is organized as follows. In~Section~\ref{sec: Gravitational Baryogenesis}, we review the standard gravitational baryogenesis mechanism and its main theoretical ingredients. In~Section~\ref{sec:Modified Gravitational Baryogenesis}, we introduce the modified interaction terms relevant for EMSG. In~Section~\ref{sec: F(R,TT) gravity}, we summarize the theoretical framework of $f(R,\mathcal T^2)$ gravity and its cosmological equations. In~Section~\ref{sec:MGravitational Baryogenesis}, we apply the modified baryogenesis framework to the representative models, including the entropy interpretation, the~Weinberg-operator decoupling temperature, and~the constraints on $\eta$, $M_\ast$, and~$c_{\mathcal T^2}$. Finally, Section~\ref{sec:Summary} presents our conclusions. Throughout this work, we use natural units, $\hbar=c=k_B=1$, and~define the reduced Planck mass as
\begin{equation} M_{Pl}=(8\pi G)^{-1/2}\simeq2.4\times10^{18}\,\mathrm{GeV}. \end{equation}

{For} notational simplicity, we write $\mathcal T^2=T_{\mu\nu}T^{\mu\nu}$.

\section{Gravitational~Baryogenesis}\label{sec: Gravitational Baryogenesis}

\subsection{Formalism}

Gravitational baryogenesis~\cite{Davoudiasl:2004gf} is a mechanism that generates baryon asymmetry while maintaining thermal equilibrium, through the interaction term:
\begin{equation}\label{GB asymmetry}
    \mathcal{L}_{int} = \frac{c_R}{M^2_\ast} \ \partial_\mu(R) J_B^\mu \,.
\end{equation}
where $c_R$ is the dimensionless Wilson coefficients controlling the strength of this Effective Field Theory (EFT) term . In~an expanding universe, this term leads to $\mathrm{CP}$ and $\mathrm{CPT}$ violating interactions by directly coupling the derivative of the Ricci scalar, $R$, with~the baryon number (B) current $J_B^\mu$. This interaction can also be applied to a lepton (L) current or any current that results in a net {B}$-${L} charge in equilibrium~\cite{Davoudiasl:2004gf}. The~parameter $M^2_\ast$ represents the cutoff scale of the effective theory, and~it is hypothesized that this interaction originates from a higher-dimensional operator in supergravity theories~\cite{Kugo:1982mr,Kugo:1983mv}.

Furthermore, as~suggested in~\cite{Davoudiasl:2004gf}, such an operator may emerge within a low-energy effective field theory of quantum gravity, particularly if the fundamental scale $M_*$ is of the order of the reduced Planck scale $M_{Pl}$. This mechanism requires baryon number-violating processes to be in thermal equilibrium to produce the asymmetry.
The key feature of this mechanism is that CP violation arises from a gravitational interaction~\cite{Davoudiasl:2004gf}. As~the universe expands, microscopic CP violation is amplified by a shift in baryon energy, leading to a dynamic violation of CPT that alters the relative energies of particles and antiparticles. As~a result, a~CP-conserving but baryon number-violating interaction in equilibrium can generate the asymmetry, which is enhanced by the term \eqref{GB asymmetry} and becomes fixed when the baryon number-violating interaction~decouples.

Hence, the~asymmetry resulting from \eqref{GB asymmetry} is derived as follows: in a flat FLRW Universe, for~a spatially constant $R$, the~interaction \eqref{GB asymmetry} induces different energy contributions for particles and antiparticles, thereby altering their thermal equilibrium distributions, in~the following way
\begin{equation}
    \frac{1}{M^2_\ast} (\partial_\mu R) J_B^\mu = \frac{1}{M^2_\ast} \dot{R}(n_B-n_{\Bar{B}}) \,.
\end{equation}

Here, we use $J_B^0 = n_B - n_{\Bar{B}}$, where $n_B$ and $n_{\Bar{B}}$ represent the baryon and anti-baryon number densities, respectively. This contribution introduces a shift in the energy of baryons, approximately ${2\dot{R}}/{M^2_\ast}$ relative to anti-baryons, leading to CPT symmetry violation. As~a result, the~chemical potential for baryons is given by $\mu_{B} = - \mu_{\Bar{B}} =  {\dot{R}}/{M^2_\ast}$. Under~thermal equilibrium conditions, in~the regime where $T \gg m$, the~net baryon number density is expressed as~\cite{Kolb:1990vq}:
\begin{eqnarray}\label{net baryon}
    n_b = \frac{g_b}{2\pi^2} \int d^3p \left[ \frac{1}{e^{ \frac{(p - \mu_B)}{T}} + 1}
    -  
   \frac{1}{e^{\frac{(p + \mu_B)}{T}} + 1} \right] \, ,
\end{eqnarray}
where $g_b$ is the number of internal degrees of freedom of baryons. In~an homogeneous and isotropic Universe, this reduces to
\begin{equation}
    n_b = \frac{g_b T^3}{6\pi^2} \left[ \pi^2 \left(\frac{\mu_B}{T} \right) + \left(\frac{\mu_B }{T}\right)^3 \right] \, .
\end{equation}

The conservation of entropy allows the number of a given particle species within a comoving volume to be expressed as the ratio of the particle number density to the entropy density. Formally, this relationship can be written as
\begin{equation}
    N_i \equiv \frac{n_i}{s}\, ,
\end{equation}
where
\begin{equation}\label{entropy}
    s= \frac{2\pi^2}{45}g_{\ast s}(T) T^3 \, ,
\end{equation}
and $g_{\ast s}(T)$ is given by
\begin{equation}
    g_{\ast s}(T) = \sum_{i={\rm bosons}} g_i \left(\frac{T_i}{T} \right)^3 + \frac{7}{8}\sum_{i={\rm fermions}} g_i \left(\frac{T_i}{T}\right)^3\, .
\end{equation}

By considering that the asymmetry is generated in thermal equilibrium, we can reasonably assume that $ g_{\ast s}=g_{\ast} $, where $ g_{\ast} $ denotes the total effective number of relativistic degrees of freedom~\cite{Kolb:1990vq} with $g_\ast=106.75$ for relativistic particles ($T\gg 100 \ \text{GeV}$) in the SM.
Therefore, if~particles are neither produced nor destroyed, as~is the case after baryogenesis has concluded, the~number density $n_i$ scales with the inverse cube of the scale factor, $n_i \propto a^{-3}$, and~$N_i$ constant. This provides a well-defined parameter for observational~purposes.

In an expanding Universe, considering the relativistic regime ($T \gg m$) and $T \gg \mu_B$, using Equations~\eqref{net baryon} and  \eqref{entropy} the asymmetry produced by the interaction given in Equation~\eqref{GB asymmetry} at the temperature where the baryon number violation processes decouple, $T_D$, is given by~\cite{Davoudiasl:2004gf}
\begin{equation}\label{asymmetry R}
    \frac{n_b}{s} \simeq  \frac{15g_b}{4\pi^2 g_{\ast}} \frac{\dot{R}}{M^2_\ast T} \Bigg|_{T_D} \, ,
\end{equation}
where $\dot{R}$ is determined based on the cosmological model used. In~essence, the~coupling term in \eqref{GB asymmetry} modifies the thermal equilibrium distribution and the chemical potential, driving the universe toward a nonzero equilibrium B-asymmetry (or L-asymmetry) through B-violating (or L-violating) interactions.
This mechanism bears a resemblance with spontaneous baryogenesis~\cite{Cohen:1987vi,Cohen:1988kt}, wherein gravitational baryogenesis involves the Ricci scalar of a gravitational background assuming the role typically played by an axion in spontaneous~baryogenesis.

In spontaneous baryogenesis, the~underlying symmetry of the theory, which ensures the conservation of total baryon number in the unbroken phase, is spontaneously broken. When this symmetry breaks, the~Lagrangian density acquires the following term~\cite{Cohen:1987vi,Cohen:1988kt,DeSimone:2016ofp}
\begin{equation}
    \mathcal{L} = \frac{(\partial_\mu \phi)}{f} J_B^\mu \, ,
\end{equation}
where $\phi$ is typically a pseudo-Nambu--Goldstone boson associated with an approximate symmetry related to the baryon number and $f$ is a parameter associated with the symmetry breaking scale. Assuming a spatially homogeneous field, $\phi \equiv \phi(t)$, in~the same way as done in gravitational baryogenesis one has
\begin{equation}\label{Chemical potential SB}
     \mathcal{L} =  \frac{\dot{\phi}}{f}J^0_B =  \frac{\dot{\phi}}{f} (n_B-n_{\Bar{B}}) \, .
\end{equation}

However, in~spite of the similarity between the spontaneous baryogenesis and gravitational baryogenesis, the~distinctions between the two primarily lie in the behavior of both entities, the~scalar field $\phi$ and the Ricci scalar. For~spontaneous baryogenesis to occur, it is imperative that the scalar field $\phi$ engaged in the process of spontaneous symmetry breaking undergoes a uniform evolution. This evolution must be such that it progresses homogeneously in one direction while contrasting with the evolution in the other direction to bring about the desired asymmetry. Moreover, a~spatially consistent representation is also essential. These requirements are fulfilled inherently in gravitational baryogenesis due to the homogeneity of the~Universe.

One challenge with the scalar field in spontaneous baryogenesis is that its minimum occurs where the average of $\dot{\phi}$ equals zero, driving the generated asymmetry to zero and complicating perturbations around this minimum. In~contrast, the~Ricci scalar, with~a nonzero mean value, $\dot{R} \sim H^3$, avoids this issue.
Additionally, as~with gravitational baryogenesis, it is possible to identify ${\dot{\phi}}/{f}$ as a chemical potential in Equation~\eqref{Chemical potential SB}. However, as~noted in~\cite{Arbuzova:2016qfh,Dasgupta:2018eha}, this interpretation is complex due to the specific representation chosen for the fermionic fields. Therefore, the~assumption that including the term $({\dot{\phi}}/{f})J^0_B \sim ({\dot{\phi}}/{f})n_b$ in the Lagrangian density of spontaneous baryogenesis leads to an energy difference between particles and antiparticles, effectively acting as a chemical potential that generates baryon asymmetry, should be approached with caution and careful~consideration.

\subsection{Instabilities}

While gravitational baryogenesis offers significant advantages, potential instabilities must be carefully considered. As~noted in~\cite{Arbuzova:2016cem} and~\cite{Arbuzova:2017vdj}, certain models incorporating the CPT-violating term \eqref{GB asymmetry} in the gravitational action and a matter Lagrangian that includes bosonic or fermionic components can lead to instabilities, rendering the cosmological model untenable.
Recently, the~authors who identified these instabilities proposed a possible solution in~\cite{Arbuzova:2023rri}. Their solution involves modifying the theory of gravity by adding a quadratic Ricci scalar term to the gravitational action. Such modifications can prevent these instabilities, supporting the argument for introducing modified gravity in the context of gravitational baryogenesis. Although~these instabilities can arise in specific constructions, their relevance is often tied to the ultraviolet completion of the theory or to the details of the fermionic sector. Since the present work employs a macroscopic, thermodynamic description of the matter sector during the early-Universe epoch relevant for baryogenesis, these instabilities are not expected to play a role at the scales and in the regime considered~here.

By employing a thermodynamic approach to $\mathcal{L}_m$, a~stable cosmological model can be achieved without the presence of such instabilities. The~inclusion of the CP-violating term~\eqref{GB asymmetry} in the gravitational action is not the focus of the standard treatment; rather, it is typically considered as an effective interaction that arises from higher-dimensional operators in quantum gravity or supergravity theories. Within~the EFT framework adopted here, this term is treated as a perturbative source for the chemical potential, and~its back-reaction on the gravitational dynamics is subdominant, as~ensured by the hierarchy \mbox{$M_\ast > T_D$}. Therefore, this issue will not be addressed in this~work.

In this study, gravitational baryogenesis is described without incorporating the specific matter sectors containing bosons or fermions done in~\cite{Arbuzova:2016cem,Arbuzova:2017vdj}. Instead, a~macroscopic description of the matter Lagrangian is used. The~gravitational action will not include the interaction term from Equation~\eqref{GB asymmetry}, which will be treated~separately.

\section{Generalized Gravitational~Baryogenesis}\label{sec:Modified Gravitational Baryogenesis}

The concept of generalizing gravitational baryogenesis emerged soon after the original formulation was introduced~\cite{Li:2004hh}. The~first modification, also known as generalized gravitational baryogenesis, involved coupling $\partial_\mu f(R)$ with $J^\mu$, where $f(R) \sim \ln(R)$, leading to the interaction term
\begin{equation}
    \mathcal{L}_{int} = -\alpha \frac{\partial_\mu R}{R} J^\mu \,.
\end{equation}

{This} operator is of a type expected to arise when heavy particles or extra dimensions are integrated out~\cite{Carroll:2003wy, Mukohyama:2003nw}, with~$\alpha$ representing the strength of this new interaction in the effective~theory.

Building on this work, a~modification to gravitational baryogenesis was explored by employing $f(R)$ gravity to describe the gravitational interaction, which subsequently influenced the outcomes of gravitational baryogenesis~\cite{Lambiase:2006dq}. These developments sparked an emerging interest in studying gravitational baryogenesis within the context of modified theories of gravity, where alternative CP-violating terms, beyond~the standard term \eqref{GB asymmetry}, were frequently considered.
In this work, we propose the introduction of specific modified terms within the framework of these theories, categorizing them into two distinct~classes.

From a phenomenological standpoint, modifying gravitational baryogenesis can address several challenges inherent in the original formulation. One key issue is generating a non-vanishing asymmetry during the radiation-dominated epoch characterized by $w = 1/3$, although~this concern can be circumvented by considering $w \approx 1/3$~\cite{Davoudiasl:2004gf}. Furthermore, scenarios that produce either an insufficient or excessive asymmetry may benefit from modified gravitational baryogenesis. This approach may also help resolve inherent issues, such as the aforementioned instabilities.
In this context, we propose two approaches to modify and expand gravitational baryogenesis {{(}We do not consider methods involving an axial vector current for $J^\mu$~\cite{Lambiase:2006md,Mukhopadhyay:2005gb,Mukhopadhyay:2007vca,Sinha:2007uh} as modifications to gravitational baryogenesis, but~rather as independent mechanisms{)}}, which can be applied separately or in~combination.

\subsection{Modified Gravitational~Theories}

The first approach involves adopting alternative descriptions of gravity, leading to different field equations, cosmological models, or~frameworks. These alternatives result in distinct outcomes for the original gravitational baryogenesis scenario. This is exemplified by the modified $f(R)$ gravity case presented in~\cite{Lambiase:2006dq}, and~further explored in well-constructed examples of cosmological framework modifications~\cite{Odintsov:2016apy, Luciano:2022knb, Luciano:2022ely}.

The motivation for altering the gravitational and cosmological framework can be linked to addressing current issues, such as the lack of a solid theoretical foundation for the accelerated expansion of the universe, cosmological tensions, singularities in GR (e.g., in~black hole interiors and cosmological solutions), the~lithium problem~\cite{Fields:2011zzb}, and~even the study of quantum gravity theories. Employing such frameworks enables a more comprehensive exploration of these alternative~approaches.

\subsection{Different Couplings with $J^\mu$}

The second approach involves generalizing the CP-violating interaction itself. Rather than coupling only the derivative of the Ricci scalar with the baryon current, other quantities can be considered for this coupling. This modification is linked to the specific modified theory of gravity being utilized and is introduced in a similar manner to spontaneous baryogenesis~\cite{Cohen:1987vi,Cohen:1988kt} or quintessential baryogenesis~\cite{DeFelice:2002ir}. The~initial generalization in this context is to consider $f(R)$ instead of $R$, leading to a term of the form:
\begin{equation}
     \frac{1}{M^2_\ast}\int d^4x \sqrt{-g} \ \partial_\mu(f(R)) J_B^\mu  \, .
\end{equation}

The first class of modified terms we consider involves altering the term \eqref{GB asymmetry} by substituting the Ricci scalar with an alternative geometric quantity.
This category includes modified gravity theories that utilize different geometric constructs to describe gravity. In~these frameworks, it is natural to modify the term \eqref{GB asymmetry} to incorporate the specific geometric quantities relevant to the particular theory, as~the Ricci scalar may not be the central term in the geometric description of gravity. This approach is classified under geometric modifications. For~example, in~extended teleparallel theories of gravity~\cite{Bhattacharjee:2020jfk,Oikonomou:2016jjh}, the~CP-violating term can be constructed as follows:
\begin{equation}
     \frac{1}{M^2_\ast}\int d^4x \sqrt{-g} \ \partial_\mu(-\mathbf{T}) J_B^\mu  \, ,
\end{equation}
where $\mathbf{T}$ is the torsion scalar leading to the general term
\begin{equation}
     \frac{1}{M^2_\ast}\int d^4x \sqrt{-g} \ \partial_\mu(-f(\mathbf{T})) J_B^\mu  \, ,
\end{equation}

{This} can analogously be applied to $f(Q)$ gravity, where $Q$ is the~non-metricity.

Other robust and high order theories of gravity like Gauss--Bonnet gravity~\cite{ Glavan:2019inb, Fernandes:2022zrq}, generalized Einsteinian cubic gravity~\cite{Bueno:2016xff,Erices:2019mkd} are also a possible path to modified gravitational baryogenesis~\cite{Odintsov:2016hgc,Bhattacharjee:2021jwm}. 

We also include in this group special cases that are done in the context of black holes~\cite{Hamada:2016jnq,Smyth:2021lkn} that typically evolve a term with the form
\begin{equation}
    \partial_\alpha(R^{\mu \nu \rho \sigma} R_{\mu \nu \rho \sigma}) J^{\alpha} \, ,
\end{equation}
where $R_{\mu \nu \rho \sigma}$ is the Riemann tensor and $R^{\mu \nu \rho \sigma} R_{\mu \nu \rho \sigma}$ is the Kretschmann~scalar.

The second class of modifications involves theories of gravity that incorporate matter sectors into the gravitational action with a nonminimal coupling (NMC) to geometry. Examples of such theories include $f(R,\mathcal{L}_m)$ gravity~\cite{Harko:2010mv} and $f(R,T)$ gravity~\cite{Harko:2011kv}  {{(}Gravitational baryogenesis within these frameworks has been studied in~\cite{Jaybhaye:2023lgr} for $f(R,\mathcal{L}_m)$ gravity and in~\cite{Baffou:2018hpe, Nozari:2018ift, Sahoo:2019pat} for $f(R,T)$ gravity{)}}.

In these theories, we propose that analogous terms to \eqref{GB asymmetry} can be constructed using the matter components $\mathcal{L}_m$, $T$, and~others. This idea is supported by the fact that theories with nonminimal couplings between matter and geometry can exhibit gravitationally-induced particle production, as~described in~\cite{Harko:2014pqa,Harko:2015pma,Pinto:2022tlu,Pinto:2023tof,Pinto:2023phl}  {{(}This phenomenon arises from the non-conservation of the energy-momentum tensor, which generally implies a violation of the Equivalence Principle. This violation is a known issue in these theories and is also observed in some dark energy and dark matter models. For~instance, observational data from the Abell Cluster A586 suggests an interaction between dark matter and dark energy, which may imply a violation of the Equivalence Principle~\cite{Bertolami:2007zm}. Furthermore, these theories often predict non-geodesic motion for particles under forces orthogonal to their four-velocity{)}}.

The possibility of gravitationally-induced particle production provides a strong rationale for exploring the direct coupling of matter-related quantities with the baryonic current. However, it is important to clarify that here gravitational particle production is not directly used to generate baryonic asymmetry. Instead, this mechanism, which arises from the non-conservation of the energy-momentum tensor in theories with geometry--matter couplings, can be integrated with gravitational baryogenesis to achieve baryonic asymmetry. This integration requires careful consideration, such as ensuring entropy conservation after the baryogenesis epoch and developing a method to produce more particles than antiparticles through some B-violating process, which may or may not be gravitationally~created.

In gravitational baryogenesis, despite the term ``gravitational'', gravity does not directly cause the B (or B-L) violation. Instead, gravity influences the net baryon current through the background, and~the baryon asymmetry results from an existing B-violating process. The~interaction term from Equation~\eqref{GB asymmetry} enhances the baryon asymmetry to its observed value, which remains fixed once these B-violating processes decouple. Therefore, a~mechanism that uses gravitational particle production to produce baryonic asymmetry, coupled with a B-violation process potentially originating from gravitational effects, and~further amplifies this asymmetry through gravitational baryogenesis, represents a primarily gravitational means of achieving baryogenesis. This holds true even when supplementary mechanisms are considered for generating the asymmetry. A~similar concept was explored in~\cite{Lima:2016cbh}, though~this work used GR for the gravitational description, leading to a different justification for gravitational particle production compared to the approach described here. More recently, in~\cite{Flores:2024lzv}, a~comparable mechanism was proposed, integrating Cosmological Gravitational Particle Production (CGPP) \cite{Kolb:2023ydq}, which combines cosmic inflation, quantum field theory, general relativity, and~particle properties to produce particles, with~baryogenesis from Grand-Unification Theories (GUT).

{ A first implementation of this idea was developed in our previous work~\cite{Pereira:2025flo}, where gravitational baryogenesis was investigated in both GR and $f(R,\mathcal{T}^2)$ gravity by coupling the baryonic current directly to the derivative of the energy-momentum-squared scalar $\mathcal{T}^2$. That analysis established $\mathcal{T}^2$ as a matter-sector source for gravitational baryogenesis and examined the associated cutoff scale and BBN constraints. The~present work retains the same underlying motivation, namely that the high-density matter sector can provide a time-dependent source for the baryon asymmetry, but~extends the construction in a substantive way. Instead of restricting the interaction to the derivative of the bare scalar $\mathcal{T}^2$, we consider derivative couplings involving the functional dependence on $\mathcal{T}^2$ as well as the complete gravitational function $f(R,\mathcal{T}^2)$. We also extend the cosmological analysis to a broader set of power-law models and consistently incorporate the modified radiation-era evolution, the~associated entropy production or dilution, and~a decoupling temperature determined by an explicit $B-L$-violating process. Thus, the~interaction introduced in Ref.~\cite{Pereira:2025flo} provides the starting point for the generalized baryogenesis couplings considered~here:}
\begin{equation}\label{eq:T2 original}
    \mathcal{L}_{\text{int}}=\frac{c_{\mathcal{T}^2}}{M_\ast^8} {\partial_\mu \mathcal{T}^2} J^\mu\,,
\end{equation}
{ where $c_{\mathcal{T}^2}$ is a dimensionless coupling constant. Motivated by this direct $\mathcal{T}^2$ coupling, we generalize the interaction by allowing the baryonic current to couple to derivatives of the corresponding matter-sector function and of the full gravitational function, leading to the interaction term having the form}
\begin{equation}
    \mathcal{L}_{\text{int}}= \frac{c_{{\mathcal{T}^2}}}{M_{\ast}^{2}} \partial_\mu \left(f\left(\mathcal{T}^2\right)\right) J^\mu\,,
\end{equation}
\begin{equation}
    \mathcal{L}_{\text{int}}= \frac{c_{{\mathcal{T}^2}}}{M_{\ast}^{2}} \partial_\mu f(R,\mathcal{T}^2) J^\mu\,,
\end{equation}
that give the asymmetries for a baryonic current
\begin{equation}\label{eq:fT2 term}
    \frac{n_b}{s} = \frac{15 g_bc_{\mathcal{T}^2}}{4\pi^2 g_\ast}\frac{ \partial_tf\left(\mathcal{T}^2\right) \, }{M_\ast^2 T_D}\,,
\end{equation}
\begin{equation}\label{eq:fRT2 term}
    \frac{n_b}{s} = \frac{15 g_bc_{\mathcal{T}^2}}{4\pi^2 g_\ast}\frac{ \partial_t \left(f\left(R,\mathcal{T}^2\right)\right) }{M_\ast^2 T_D}\,.
\end{equation}

If, instead, one considers a $B\!-\!L$ current, the~resulting asymmetry is given by
\begin{equation}\label{eq:general term T2}
    \frac{n_{B-\!L}}{s} \simeq \frac{15\,\hat{g}\,c_{\mathcal{T}^2}}{4\pi^2\,g_{\ast }}\,
    \frac{\partial_t f\!\left(\mathcal{T}^2\right)}{M_\ast^2\,T_D}\,,
\end{equation}
and, for~a generic interaction $f(R,\mathcal{T}^2)$, by~\begin{equation}\label{eq:general term RT2}
    \frac{n_{B-\!L}}{s} \simeq \frac{15\,\hat{g}\,c_{\mathcal{T}^2}}{4\pi^2\,g_{\ast }}\,
    \frac{\partial_t f\!\left(R,\mathcal{T}^2\right)}{M_\ast^2\,T_D}\,,
\end{equation}
where $\hat{g}\equiv \sum_i g_i q_i^2$ and $q_i\equiv(B\!-\!L)_i$. For~the Standard Model (SM)
with three fermion families and no right-handed neutrinos, one finds $\hat{g}=13$ (or $\hat{g}=16$
upon the inclusion of three right-handed neutrinos). Since the observed baryon asymmetry is generated
from a primordial $B\!-\!L$ asymmetry through electroweak sphaleron processes, sphaleron equilibrium
relates the baryon-to-entropy ratio to the $B\!-\!L$ asymmetry as~\cite{Bodeker:2020ghk,Harvey:1990qw}
\begin{equation}\label{eq:BL to B}
    \frac{n_b}{s}= k_s\,\frac{n_{B\!-\!L}}{s}\,,
\end{equation}
with
\begin{equation}
    k_s=\frac{8N_f+4N_H}{22N_f+13N_H}\,,
\end{equation}
where $N_f$ denotes the number of fermion families and $N_H$ the number of Higgs doublets, yielding
$k_s=\frac{28}{79}$ for the~SM.

Our goal with the generalized interaction term is to assess the impact of such couplings on baryogenesis by expanding the interaction term~\eqref{eq:T2 original}. In~this way, one can directly probe how geometry--matter interactions affect the generation of the baryon asymmetry and further explore the interface between gravitational dynamics and particle-physics processes in the early~Universe.

We emphasize that, in~the present work, the~baryogenesis operators are treated within the spectator approximation. As~recently discussed in Ref.~\cite{Pereira:2026djd}, going beyond this approximation requires including the derivative baryogenesis operator directly in the gravitational action and varying it with respect to the metric, which generically produces additional curvature--matter-coupling terms and modifies the background field equations. Here we do not include this backreaction. Instead, the~cosmological background is determined solely by the $f(R,\mathcal T^2)$ field equations, while the derivative interaction is used only as an effective source for the chemical potential that generates the baryon or $B-L$ asymmetry. This approximation is consistent provided the baryogenesis operator remains subleading in the gravitational dynamics, as~ensured by the EFT hierarchy $M_\ast>T_D$ and by the perturbative choice $|c_{\mathcal T^2}|\leq1$.

\section{\boldmath{$f(R,\mathcal{T}^2)$} Gravity}\label{sec: F(R,TT) gravity}


One of the main reasons for exploring alternative theories of gravity beyond GR is the observed accelerated expansion of the universe. The~most common explanation for this acceleration is dark energy, which is thought to drive the expansion by exerting a negative pressure that counteracts the gravitational attraction, resulting in an accelerated expansion of the universe.
The leading candidate for dark energy is the cosmological constant, which represents a constant energy density added to Einstein's equations through the term $2\Lambda$. This term is included in the Einstein--Hilbert action~\cite{Hilbert1915} to produce the observed acceleration. However, this solution faces several challenges. The~theoretical origin of dark energy remains unclear, and~observational data do not perfectly align with the theoretical description of the cosmological constant as vacuum energy. These issues have led to the search for more sophisticated solutions to the problem of the late-time~acceleration.

One notable alternative is the $f(R)$ theory of gravity, first introduced by A.H. Buchdahl~\cite{Buchdahl:1970ynr}, where $f$ is a function of the Ricci scalar $R$. This theory has been thoroughly examined, with~detailed discussions available in sources like~\cite{DeFelice:2010aj,Sotiriou:2008rp}. In~addition, more complex gravitational theories have emerged and gained interest, such as Teleparallel Gravity~\cite{Bahamonde:2021gfp}, Horndeski/Galileon theories~\cite{Deffayet:2013lga}, and~$f(R, \mathcal{L}_m)$ gravity~\cite{Harko:2010mv}, $f(R,T)$ gravity~\cite{Harko:2011kv}, and~$f(R, \mathcal{T}^2)$ gravity~\cite{Katirci:2013okf}.
These theories were initially developed to address the late-time acceleration problem, but~they also show promise in tackling issues related to spacetime singularities that occur within the framework of GR. Since GR alone is insufficient due to the anticipated effects of quantum gravity, theories like $f(R, \mathcal{T}^2)$ provide a compelling~alternative.

\subsection{Action and Field~Equations}

The action for $f(R,\mathcal{T}^2)$  gravity reads
\begin{equation}\label{actionsf(R,T2)}
	S=\frac{1}{2\kappa}\int d^4x \sqrt{-g} f(R,\mathcal{T}^2) +\int d^4x \sqrt{-g} \mathcal{L}_m \, ,
\end{equation}
where $\kappa = 8\pi G = M_{Pl}^{-2}$ and $f(R,\mathcal{T}^2)$ is a well behaved function of the Ricci scalar and $\mathcal{T}^2$.

We define the energy-momentum tensor of the matter fields as
\begin{equation}\label{Stresse-energy tensor}
	T_{\mu\nu} = -\frac{2}{\sqrt{-g}}\frac{\delta (\sqrt{-g}\mathcal{L}_M)}{\delta g^{\mu\nu}}\, , 
\end{equation}
and imposing that $\mathcal{L}_m$ depends solely on the metric components and not on their derivatives we obtain
\begin{equation}
	T_{\mu\nu} = g_{\mu\nu} \mathcal{L}_m - \frac{\partial \mathcal{L}_m }{\partial g^{\mu\nu} } \, .
\end{equation}

Varying the action \eqref{actionsf(R,T2)} with respect to the metric yields the field equations
\begin{equation}\label{FE actionsf(R,T2)}
	f_{,R} R_{\mu\nu} -\frac{1}{2}g_{\mu\nu}f +  (g_{\mu\nu} \nabla^\alpha \nabla_\alpha -\nabla_\mu \nabla_\nu)f_{,R} = \kappa (T_{\mu\nu} - \frac{1}{\kappa}f_{\mathcal{T}^2} \theta_{\mu\nu}) \, ,
\end{equation}
where subscripts denote differentiation with respect to the respective quantity and $\theta_{\mu\nu}$ is defined as
\begin{eqnarray}
	\theta_{\mu\nu} \equiv \frac{\delta T^{\alpha\beta} T_{\alpha\beta}}{\delta g^{\mu\nu}} 
	& = &
	-2\mathcal{L}_m(T_{\mu\nu} - \frac{1}{2}g_{\mu\nu}T) - TT_{\mu\nu}  
	\nonumber  \\ 
	&& + 2T^{\alpha}_\mu T_{\nu\alpha} 
	- 4T^{\alpha\beta} \frac{\partial^2 \mathcal{L}_m}{\partial g^{\mu\nu} \partial g^{\alpha\beta}} .
\end{eqnarray}

For the $f(R,\mathcal{T}^2)$ form, we considered a particular class of models given by
\begin{equation}\label{model f(R,TT)}
	f(R,\mathcal{T}^2) = R + \eta M_{Pl}^{2-8n} \left(\mathcal{T}^2\right)^n \, ,
\end{equation}
that is dubbed as energy-momentum powered gravity, which is a special case of $f(R,\mathcal{T}^2)$ gravity. In~this context, $\eta$ is a dimensionless constant that is responsible for determining the coupling strength of the scalar $\mathcal{T}^2$ and $n$ is the parameter that specifies the model. We define $\eta' \equiv \eta M_{Pl}^{2-8n}$, which will be used interchangeably with $\eta$ where appropriate, with~the understanding that $\eta'$ carries the appropriate mass dimension. For~$n=1$, the~model corresponds to EMSG~\cite{Roshan:2016mbt} and $n=1/2$ and $n=1/4$ are models that were studied in detail in the literature~\cite{Katirci:2013okf}. 

At the action level, the~structure of the model under consideration resembles GR with $\mathcal{L}_m + f(\text{matter})$, where $f(\text{matter})$ is a function depending on the matter components introduced by the modified gravity framework, which specifically, in~our case, is $f(\text{matter}) = f(\mathcal{T}^2)$. Models of this kind yield field equations of the form:
\begin{equation}\label{Field Equations model}
	R_{\mu\nu}-\frac{1}{2}g_{\mu\nu}R=\kappa T^{\text{tot}}_{\mu\nu}\,.
\end{equation}
where $T^{\text{tot}}_{\mu\nu}$ is the sum of the standard matter energy-momentum tensor $T_{\mu\nu}$ and the new part, $T^{\text{mod}}_{\mu\nu}$, that arises from $f(\text{matter})$ being variationally defined as
\begin{equation}
	T^{\text{mod}}_{\mu\nu} = - \frac{2}{\sqrt{-g}}\frac{\delta(\sqrt{-g}f(\text{matter}))}{\delta g^{\mu\nu}} \,.
\end{equation}

For the specific model considered in this work, $T^{\text{mod}}_{\mu\nu}$ has the form
\begin{equation}\label{new T uv}
	T^{\text{mod}}_{\mu\nu} = \frac{\eta'}{2\kappa}(\mathcal{T}^2)^n g_{\mu\nu} - \frac{\eta'n}{\kappa} (\mathcal{T}^2)^{n-1}\theta_{\mu\nu} \, ,
\end{equation}
so that $T^{\text{tot}}_{\mu\nu}$ is given by
\begin{equation}
	T^{\text{tot}}_{\mu\nu}= T_{\mu\nu}+\frac{
		\eta'}{\kappa}(\mathcal{T}^2)^{n-1}\left[\frac{1}{2}
	(\mathcal{T}^2)g_{\mu\nu}-n\theta_{\mu\nu}\right] \, .
\end{equation}

{This} results in the trace of the field Equation \eqref{Field Equations model} being given by
\begin{equation}\label{FE trace}
	2\left(\mathcal{T}^2\right)^n - n\left(\mathcal{T}^2\right)^{n-1}\theta = -\frac{1}{\eta'}\left( \kappa T +R\right) \, ,
\end{equation}
where $\theta$ is the trace of $\theta_{\mu\nu}$ and $T$ the trace of the energy-momentum tensor. This equation allows to evaluate $\mathcal{T}^2$ depending on the choice of $n$.

The interpretation of this model for $f(R, \mathcal{T}^2)$ as GR with $\mathcal{L}_m^{\text{tot}}$, where $\mathcal{L}_m^{\text{tot}} = \mathcal{L}_m + f(\text{matter})$, was recently employed to demonstrate an equivalence between matter-coupled modified gravity theories and GR with nonminimal matter interaction~\cite{Akarsu:2023lre}. 
More importantly, this work demonstrated that in these types of models, where the matter sector can be interpreted as the sum of the usual matter Lagrangian and $f(\text{matter})$, the~term $\frac{\partial^2 \mathcal{L}_m}{\partial g^{\mu\nu} \partial g^{\alpha\beta}}$, arising from the variation of $f(\text{matter})$ with respect to the metric and often considered zero or neglected, is a crucial foundational aspect of these theories.
Furthermore, if~one chooses to omit this second derivative term, then the prior interpretation $\mathcal{L}_m^{\text{tot}} = \mathcal{L}_m + f(\text{matter})$ no longer represents the total matter Lagrangian of the model, as~a term resulting from the variation of $\sqrt{g} f(\text{matter})$ is disregarded. In~\cite{Akarsu:2023lre}, it was also shown that $\frac{\partial^2 \mathcal{L}_m}{\partial g^{\mu\nu} \partial g^{\alpha\beta}} \neq 0$ for $\mathcal{L}_m = p$ and $\mathcal{L}_m = -\rho$, reinforcing the argument that omitting this term limits the model's validity to the field equations, thereby requiring a compromise on the Lagrangian formulation.
Indeed, if~this term is retained, it leads to a description with a nonminimal matter interaction, similar to approaches explored in the context of dark energy and dark matter~\cite{Akarsu:2023lre}. Since the Standard Model cannot fully explain these phenomena, modified theories of gravity that extend the GR paradigm by introducing higher-order contributions of matter may play a crucial role in understanding these components of the Universe. Additionally, within~the scope of this work, the~high-energy epoch during which baryogenesis is thought to occur is inadequately described by the Standard Model, suggesting that gravitational dynamics could play a significant role in capturing its complexities. Contributions from higher-order matter terms could illuminate aspects of the baryogenesis epoch, potentially offering insights into the connections between particle physics and gravity and aiding in the ongoing quest for a quantum theory of~gravity.

In this work, we adopt the choice $\mathcal{L}_m = p$ \cite{Katirci:2013okf, Bertolami:2007gv, schutz1970perfect, Odintsov:2013iba, Bertolami:2008ab}, as~it effectively characterizes the energy properties of the primordial universe during the baryogenesis epoch. This choice provides an accurate thermodynamic model of the early universe and enables a detailed analysis of the theory's implications for gravitational baryogenesis. However, consistent with the findings of~\cite{Akarsu:2023lre}, this approach introduces a divergence at $p=0$ due to the term $\frac{\partial^2 \mathcal{L}_m}{\partial g^{\mu\nu} \partial g^{\alpha\beta}}$.
Thus, to~avoid the emergence of singularities, one can set the second derivative term to zero, as~is commonly done in the literature. To~clarify this important point further, starting from Equation~\eqref{Field Equations model}, the~twice-contracted Bianchi identity, $\nabla_\mu(R_{\mu\nu}-\frac{1}{2}g_{\mu\nu}R) = 0$, yields $\nabla_\mu(T_{\mu\nu} + T^{\text{mod}}_{\mu\nu}) = 0$, a~result that does not imply $\nabla_\mu(T_{\mu\nu}) = \nabla_\mu(T^{\text{mod}}_{\mu\nu}) = 0$. Therefore, for~accuracy, one should consider the existence of a potential nonminimal interaction between the material field and the modification sector, as~given by~\cite{Akarsu:2023lre}.
\begin{equation}
	\nabla^\mu T_{\mu\nu} = - Q_\nu = - \nabla^\mu T^{\text{mod}}_{\mu\nu} \, ,
\end{equation}
where $Q_\nu$ is the interaction kernel that governs the rate of energy transfer (for further details see~\cite{Akarsu:2023lre}). For~the model under question $Q_\nu$ is given by
\begin{equation}
	Q_\nu = \eta' \partial_\nu \left((\mathcal{T}^2)^n\right) - 2n\eta'\theta_{\mu\nu} \partial^{\mu} \left((\mathcal{T}^2)^{n-1}\right)-2n\eta'(\mathcal{T}^2)^{n-1} \nabla^\mu \left(\theta_{\mu\nu}\right) \, .
\end{equation}

By analyzing the previous equation and Equation~\eqref{new T uv} one sees that the function $f$ and the tensor $\theta_{\mu\nu}$ coexist in both equations, intertwining the EoS of the second source described by $T^{\text{mod}}_{\mu\nu}$ and $Q_\nu$, see~\cite{Akarsu:2023lre} for in-depth details. In~line with~\cite{Akarsu:2023agp}, we will employ the function $f$ to develop the form of the interaction kernel $Q_\nu$. 
Given the absence of an equation of state for the source produced by $T^{\text{mod}}_{\mu\nu}$ one has the freedom to adjust the expression for $\theta_{\mu\nu}$~\cite{Akarsu:2023lre}. To~accommodate the divergences arising from $\frac{\partial^2 \mathcal{L}_m}{\partial g^{\mu\nu} \partial g^{\alpha\beta}}$ we will set this term to zero, aligning with the existing literature. Therefore, the~expression for the tensor $\theta_{\mu\nu}$ is as follows~\cite{Akarsu:2023lre}
\begin{equation}\label{theta}
	\theta_{\mu\nu} = 
	-2\mathcal{L}_m(T_{\mu\nu} - \frac{1}{2}g_{\mu\nu}T) - TT_{\mu\nu}  + 2T^{\alpha}_\mu T_{\nu\alpha} \, .
\end{equation}

Having this defined, relative to the matter sector, we assume that it is described by a perfect fluid
\begin{equation}
	T_{\mu\nu} = (\rho + p) u_\mu u_\nu + pg_{\mu\nu} \, ,
\end{equation}
where $\rho$ is the energy density and $p$ the isotropic pressure. For~this description, the~trace of the energy-momentum tensor, the~scalar $\mathcal{T}^2$ and the tensor $\theta_{\mu\nu}$ are given by
\begin{equation}
	T= 3p-\rho \, ,
\end{equation}
\begin{equation}\label{Eq T^2}
	\mathcal{T}^2 = \rho^2 + 3p^2 \,,
\end{equation}
and
\begin{equation}
	\theta_{\mu\nu}=-(\rho^2 + 4p\rho + 3p^2) u_\mu u_\nu
\end{equation}
respectively.

\subsection{Cosmology in~EMSG}\label{sec:Cosmology}

To develop the foundational framework of gravitational baryogenesis, it is essential to provide a cosmological description of the primordial universe, particularly focusing on the radiation-dominated era, during~which the baryon asymmetry is considered to be generated. For~this purpose, we assume a flat FLRW (Friedmann--Lemaître--Robertson--Walker) metric, given by
\begin{equation}\label{metric}
	\textrm{d}s^{2}=- \textrm{d}t^{2}+a^{2}(t)\textrm{d}V^2 ,  
\end{equation}
where $t$ is the comoving proper time, $a(t)$ is the expansion scale factor and $\textrm{d}V$ is the volume element in comoving~coordinates.

Using this metric, the~modified Friedmann equations are
\begin{equation}\label{friedmann}
	H^2 =\kappa \frac{\rho }{3}+\frac{\eta'}{3}(\rho ^{2}+3p^{2})^{n-1}\left[(n- 
	\frac{1}{2})(\rho ^{2}+3p^{2})+4n\rho p\right]\, ,  
\end{equation}
where $H\equiv {\dot{a}}/{a}$ is the Hubble parameter, and~the acceleration equation
\begin{equation}\label{acceleration}
	\dot{H} + H^2=-\kappa \frac{\rho +3p}{6}-\frac{\eta' 
	}{3}(\rho ^{2}+3p^{2})^{n-1} \left[ \frac{n+1}{2}(\rho ^{2}+3p^{2})+2n\rho p%
	\right]\, ,
\end{equation}
respectively. 

Assuming that the matter fields obey a barotropic equation of state, $p = w\rho$, where $w$ is a constant, the~additional terms introduced by the scalar $\mathcal{T}^2$ are all of the form $\rho^{2n}$, each multiplied by a constant. Therefore, the cosmological equations can be recast as~\cite{Board:2017ign}
\begin{equation}\label{Fridmann T^2}
	H^2=\kappa \frac{\rho }{3}+\frac{\eta' \rho ^{2n}}{3} C_{Frd}(n,w) \, ,
\end{equation}
where $C_{Frd}$ is a function that depends on the choice of $n$ and $w$ and is given by
\begin{equation}\label{Acc T^2}
	C_{Frd}(n,w)\equiv (1+3w^{2})^{n-1}\left[(n-\frac{1}{2})(1+3w^{2})+4nw\right]\, ,
\end{equation}
and the acceleration equation becomes
\begin{equation}\label{Acel}
	\dot{H} + H^2=-\kappa \frac{1+3w}{6}\rho -\frac{\eta'
		\rho ^{2n}}{3}C_{Acc}(n,w) \, ,  
\end{equation}
where $C_{Acc}$ provides a constant based on the choice for $n$ and $w$, and~is given by
\begin{equation}
	C_{Acc}\equiv (1+3w^{2})^{n-1}\left[\frac{n+1}{2}(1+3w^{2})+2nw\right] \, .
\end{equation}

{The} term $\rho^{2n}$ present in both equations resembles quantum geometry effects in loop quantum gravity~\cite{Ashtekar:2006uz} and in braneworlds~\cite{Shtanov:2002mb}.

Finally, the~modified continuity equation is obtained by direct derivation from the Friedmann equation \eqref{friedmann} leading to
\begin{equation}\label{modified continuity}
	\dot{\rho}=-3H\rho(1+w)C_{con}(n,w) \, ,
\end{equation}
where $C_{con}(n,w)$ is the extra term originated from $\mathcal{T}^2$, and~is given by
\begin{equation}
	C_{con}(n,w) \equiv \left[\frac{\kappa+\eta'\rho^{2n-1} n(1+3w)(1+3w^2)^{n-1}}{\kappa+2\eta'\rho^{2n-1}n C_{Frd}(n,w)}\right] \, .
\end{equation}

An analysis of the cosmological equations reveals that the cosmological description is largely determined by the three functions under consideration and the chosen value of $n$. For~the baryogenesis epoch, we focus on the radiation-dominated era following inflation, thus, avoiding the issue of asymmetry washout due to~inflation. 

Importantly, for~the radiation-dominated epoch, $w=1/3$, the~modified continuity equation
takes the form
\begin{equation}\label{eq:rad_continuity_Ccon}
	\dot{\rho}=-4H\rho\,C_{\rm con}.
\end{equation}

It is now important to distinguish Equation~\eqref{eq:rad_continuity_Ccon} from the
local thermal-equilibrium relation
\begin{equation}\label{eq:rho_s_thermal}
	\rho=\frac{\pi^2}{30}g_\ast T^4,
	\qquad
	s=\frac{2\pi^2}{45}g_{\ast s}T^3.
\end{equation}

{The} relation $\rho\propto T^4$ defines the temperature of the radiation bath and does not, by~itself, imply adiabatic expansion. The~standard adiabatic radiation limit is recovered only when $C_{\rm con}=1$. For~the model $f(R,\mathcal{T}^2)=R+\eta M_{Pl}^{2-8n}\left(\mathcal{T}^2\right)^n$, the~radiation-era continuity factor can be written as
\begin{equation}\label{eq:Ccon_rad_general}
	C_{\rm con}=\frac{1+2nX}
	{1+\frac{4n(4n-1)}{3}X},
\end{equation}
where
\begin{equation}\label{eq:X_rad_general}
	X
	=
	\eta
	\left(\frac{4}{3}\right)^{n-1}
	\left(\frac{\rho}{M_{Pl}^4}\right)^{2n-1}.
\end{equation}

{Thus}, the~quantity $C_{\rm con}-1$ directly measures the departure from standard adiabatic radiation evolution. Consequently, the~standard adiabatic interpretation of $n_B/s$ requires additional care. The~quantity computed by the gravitational baryogenesis source is the asymmetry at freeze-out,
\begin{equation}
	Y_B(T_D)\equiv \left.\frac{n_B}{s}\right|_{T=T_D}.
\end{equation}

{After} the $B\!-\!L$-violating interaction decouples, the~comoving baryon
number is conserved. However, because~the matter energy-momentum tensor is not
generally conserved in $f(R,\mathcal{T}^2)$ gravity, the~comoving entropy
$S=sa^3$ need not be~conserved.

Using Equations~~\eqref{eq:rad_continuity_Ccon} and \eqref{eq:rho_s_thermal}, one
finds
\begin{equation}\label{eq:T_evolution_Ccon}
	\frac{\dot T}{T}=-H C_{\rm con}.
\end{equation}

{Since} $s\propto T^3$, the~comoving entropy evolves according to
\begin{equation}\label{eq:entropy_evolution_a}
	\frac{d\ln S}{d\ln a}=3(1-C_{\rm con}).
\end{equation}

{Therefore}, the final baryon asymmetry is related to the freeze-out value by
\begin{equation}\label{eq:YB_final_general}
	Y_B^{\rm final}
	=
	Y_B(T_D)\,\Delta_S^{-1},
\end{equation}
where
\begin{equation}\label{eq:DeltaS_general_a}
	\Delta_S
	=
	\frac{S_f}{S_D}
	=
	\exp\left[
	\int_{a_D}^{a_f}3(1-C_{\rm con})\,d\ln a
	\right].
\end{equation}

{Equivalently}, using $d\ln T/d\ln a=-C_{\rm con}$,
\begin{equation}\label{eq:DeltaS_general_T}
	\Delta_S
	=
	\exp\left[
	\int_{T_D}^{T_f}
	\frac{3(C_{\rm con}-1)}{C_{\rm con}}\,d\ln T
	\right].
\end{equation}

{Here} $T_f$ denotes the temperature at which the theory has returned to the standard adiabatic radiation regime, or~conservatively the temperature relevant for BBN. If~$C_{\rm con}\simeq1$, then $\Delta_S\simeq1$, and~the usual comparison between $Y_B(T_D)$ and the observed baryon asymmetry is justified. If~$C_{\rm con}$ differs significantly from unity, the~computed asymmetry must be interpreted as a freeze-out value and corrected by the entropy factor $\Delta_S$. 

A particularly interesting model arises when one imposes $C_{\text{con}}=1$ that recovers the standard $ \dot{\rho}=-4H\rho$ and the comoving entropy is conserved $\Delta_S=1$. Imposing this condition yields
\begin{equation}
	2nX = \frac{4n(4n-1)}{3}X\,,
\end{equation}
and requiring this result to hold independently of $X$ gives
\begin{equation}
	n=\frac{5}{8}\,.
\end{equation}

Therefore, for~this model the freeze-out asymmetry is directly equal to the
final baryon asymmetry
\begin{equation}
	Y_B^{\rm final}=Y_B(T_D).
\end{equation}

Moreover, the~value $n=5/8$ is singled out by the structure of the modified continuity equation. As~discussed by Board and Barrow~\cite{Board:2017ign}, the~continuity equation can reduce to its standard GR form for specific choices of the power $n$.

With this framework established, $w=1/3$ will be used for all models. In~this context, the~new terms associated with each cosmological equation depend only on the specific model chosen, particularly the power of $\mathcal{T}^2$. This highlights the importance of the model in shaping the cosmological framework and its parameters. Additionally, the~coupling constant $\eta$ is expected to play a key role in cosmological dynamics, significantly influencing the asymmetry produced through gravitational baryogenesis. Since the generated asymmetry largely depends on this parameter, $\eta$ can be constrained by imposing observational limits on the resulting~asymmetry.

\section{Modified Gravitational Baryogenesis in~EMSG}\label{sec:MGravitational Baryogenesis}

In this section, we examine the effects of $f(R,\mathcal{T}^2)$ gravity in the context of modified gravitational baryogenesis and constrain $\eta$ using the observational limit on baryonic asymmetry, specifically $ Y_B = (8.8\pm 0.6) \times 10^{-11} $ \cite{WMAP:2003ogi, Planck:2018vyg, Fields:2019pfx, ParticleDataGroup:2020ssz}. We examine four values of $n$:
$n=\tfrac{1}{4}$, $n=\tfrac{1}{2}$, $n=\frac{5}{8}$ and $n=1$, and~assess their impact on modified gravitational
baryogenesis (see Section~\ref{sec:Modified Gravitational Baryogenesis}). The~choices $n=\tfrac{1}{2}$ and $n=1$ have previously been investigated in the context of Ricci- and $\mathcal{T}^2$-dependent interaction terms in Ref.~\cite{Pereira:2025flo}, where they exhibited strong phenomenological performance. Accordingly, these values provide well-motivated benchmarks for analyzing the interaction terms in Equations~\eqref{eq:fT2 term} and \eqref{eq:fRT2 term}. Additionally, we will also explore the extreme cases where $n < 1/4$ and $n > 1$.

The first step in this study involves carefully selecting the values of $M_\ast$ and $T_D$, as~these are critical parameters requiring close attention. In~the original gravitational baryogenesis framework, as~noted in~\cite{Davoudiasl:2004gf}, $M_\ast$ does not necessarily need to reach the Planck scale. For~instance, in~scenarios where B-violation is characterized by the Majorana mass $M_R$ of the right-handed neutrino, which represents a soft B-violation, the~term \eqref{GB asymmetry} avoids unitarity violation up to the Planck mass, even when $M_\ast^2 = M_R M_{Pl}$~\cite{Davoudiasl:2004gf}. As~a result, the~cutoff scale for the $\partial_\mu R$ coupling can be lowered if necessary to support baryogenesis. For~the modified couplings introduced in an ad~hoc manner, the~cutoff scale will depend on various scenarios and numerical outcomes that lead to successful baryogenesis. If~the cutoff scale were sufficiently low, such derivative interactions could in principle lead to phenomenological signatures at colliders. A~dedicated collider analysis would be required, however, before~translating a value of $M_\ast$ into a direct exclusion~bound.

For the decoupling temperature, it is generally accepted in the literature that during the radiation-dominated epoch following inflation, the~decoupling temperature is approximately equal to the inflationary scale, $M_I$, where $M_I \simeq 1.6 \times 10^{16}$ GeV, representing the upper bound based on tensor mode fluctuation constraints~\cite{Planck:2018jri}. While this approximation is widely used, our study aims to examine this assumption more critically.
During slow-roll inflation, as~inflation ends, the~scalar field begins oscillating around the potential minimum, initiating the reheating phase essential for the thermalization of the Universe. Therefore, if~baryogenesis is assumed to occur during the radiation era post-inflation, the~decoupling temperature should satisfy the condition $T_D < T_{RD} < M_I$, where $T_{RD}$ is the temperature at which the Universe becomes radiation-dominated, or~equivalently, the~reheating temperature following inflation. The~specific value of $T_{RD}$ depends on the chosen inflationary model (see comprehensive reviews and analyses in~\cite{Bassett:2005xm, Allahverdi:2010xz, Lozanov:2019jxc, Amin:2014eta}), and~while it can theoretically reach up to $\sim$$10^{16}$ GeV, similar to the GUT scale, this is often considered unrealistic. Consequently, the~assumption $T_D \approx M_I$ is problematic, as~it would require $T_{RD}$ to be nearly equal to $M_I$, a~scenario typically not~realized.

Furthermore, the~decoupling temperature should not be regarded as a free parameter. Since the baryogenesis mechanism relies on the presence of a $B$ or $B\!-\!L$ violating process, the~decoupling temperature $T_D$ is fixed by the condition $\Gamma(T_D)=H(T_D)$. Consequently, $T_D$ is determined by both the microphysical rate associated with the chosen $B$ or $B\!-\!L$ violating mechanism and the cosmological expansion history implied by the underlying gravitational theory. In~this work, we will adopt a concrete source of $B\!-\!L$ violation, namely the Weinberg
operator~\cite{Weinberg:1979sa,Turner:2018mwh,Pascoli:2018cqk},
\begin{equation}\label{eq:Weinberg}
\mathcal{L}_\text{W}=-\frac{\lambda_{\alpha\beta}}{\Lambda}\,
\ell_{\alpha L}^{i}\,\varepsilon^{ij}\,H^{j}\,C\,\ell_{\beta L}^{k}\,\varepsilon^{kl}\,H^{l}
+\text{h.c.}\,,
\end{equation}
where $\ell_L=(\nu_L,l_L)^T$ denotes the lepton doublet in the $SU(2)_L$ gauge space,
$\lambda_{\alpha\beta}=\lambda_{\beta\alpha}$ are effective Yukawa couplings with flavour indices
$\alpha,\beta=e,\mu,\tau$, and~$C$ is the charge-conjugation matrix. In~the early Universe,
this operator mediates $\Delta L=2$ scattering processes such as $LL\leftrightarrow HH$ (as well
as the corresponding crossed channels). The~thermally averaged interaction rate associated with these
processes can be expressed directly in terms of the light-neutrino masses as~\cite{Weinberg:1979sa,Turner:2018mwh,Pascoli:2018cqk}
\begin{equation}
\label{eq:GammaW}
\Gamma_W(T)=\frac{3}{4\pi^{3}}\,\frac{\bar{m}_{\nu}^{2}}{v_H^{4}}\,T^{3}\,,
\end{equation}
where $\bar m_\nu^2 \equiv \sum_{i=1}^3 m_{\nu_i}^2$ is the sum of the squared light-neutrino mass
eigenvalues. After~electroweak symmetry breaking, the~Weinberg operator generates the Majorana mass
matrix $(m_\nu)_{\alpha\beta}=\lambda_{\alpha\beta}v_H^2/\Lambda$, with~$v_H=246~\mathrm{GeV}$
the Higgs vacuum expectation value (VEV).
For the neutrino sector, we will adopt a minimal benchmark for the light-neutrino spectrum, namely normal ordering with negligible lightest mass, in~which case oscillation data fix
\begin{equation}
\bar m_\nu^{\,2}\simeq \Delta m_{21}^2+\Delta m_{31}^2 \simeq 2.6\times10^{-3}\,\mathrm{eV}^2\,,
\end{equation}
within current data~\cite{ParticleDataGroup:2024cfk,Esteban:2024eli}. 

The use of the Weinberg operator as an effective interaction also imposes an EFT consistency requirement on the decoupling temperature. In~the normalization of \mbox{Equation~\eqref{eq:Weinberg}}, the~light-neutrino mass matrix is generated after electroweak symmetry breaking according to $(m_\nu)_{\alpha\beta}=\lambda_{\alpha\beta}v_H^2/\Lambda$. Therefore, for~order-one perturbative Wilson coefficients, the~characteristic scale associated with the Weinberg operator can be estimated as
\begin{equation}\label{eq:LambdaW_EFT}
    \Lambda_W^{\rm eff} \equiv \frac{v_H^2}{\bar m_\nu}= \frac{v_H^2}{\sqrt{\bar m_\nu^{\,2}}}\simeq 1.2\times 10^{15}\,\mathrm{GeV},
\end{equation}
where we have used $v_H=246\,\mathrm{GeV}$ and $\bar m_\nu^{\,2}\simeq2.6\times10^{-3}\,\mathrm{eV}^2$. The~EFT description based on Equation~\eqref{eq:Weinberg} is, thus, self consistent provided
\begin{equation}\label{eq:Weinberg_EFT_condition}
    T_D < \Lambda_W^{\rm eff}.
\end{equation}

{As} shown below, the~decoupling temperatures obtained for all benchmark models considered in this work satisfy this~condition.

\subsection{$f(\mathcal{T}^2)$ Interaction~Term}\label{subsec:f(T2)coup}

\subsubsection{{Model $n=1/4$}}\label{fT2 n 1/4}

The model with $n=1/4$ presents interesting cosmological equations. The~modified Friedmann equation is equal to the standard Friedmann equation in GR
\begin{equation}\label{Friedmann 1/4}
    H^2 = \frac{\rho}{3M_{Pl}^2} \, ,
\end{equation}
and the modified acceleration equation is
\begin{equation}\label{Acc 1/4}
    \dot{H} + 2H^2 =  - \frac{\eta}{192^{1/4}}\rho^{1/2} \, .
\end{equation}

Another noteworthy aspect of this model is the fact that the coupling constant in this scenario is dimensionless and $\kappa$ couples to this parameter. By~contracting both equations, we derive the following differential equation
\begin{equation}
    \dot{H} + 2H^2 + z(\eta) H = 0 \, , 
\end{equation}
with
\begin{equation}
    z(\eta) = \frac{\eta \sqrt{3} M_{Pl}}{192^{1/4}} \, ,
\end{equation}
and that yields the solution
\begin{equation}\label{H solution n=1/4}
    H(t) = \frac{z}{e^{zt}-2} \, .
\end{equation}

Substituting Equation~\eqref{H solution n=1/4} into Equation~\eqref{Friedmann 1/4} we obtain the time dependent expression for the energy density
\begin{equation}\label{rho(t) n=1/4}
    \rho(t) = 3M_{Pl}^2\left(\frac{z}{e^{zt}-2}\right)^2 \, ,
\end{equation}
allowing to relate the time of decoupling with the temperature of decoupling by using the equation that relates the total radiation density with the energy of all relativistic species~\cite{Kolb:1990vq}
\begin{equation}\label{rho vs temperature}
    \rho(t)=\frac{\pi^2}{30}g_\ast T^4 \, ,
\end{equation}
and solving for $t_D$. Substituting this result into Equation~\eqref{rho(t) n=1/4} gives the relation
\begin{equation}\label{result 1}
    \left(\frac{z}{e^{zt_D}-2}\right)^2=\frac{\pi^2g_\ast}{90M_{Pl}^2}T^4_D \, ,
\end{equation}
and
\begin{equation}\label{result 2}
    e^{zt_D}=\frac{3z\sqrt{10}M_{Pl}}{\pi g^{1/2}_\ast T^2_D} + 2  \, ,
\end{equation}

Using Equation~\eqref{Eq T^2} for this case one has
\begin{equation}
    \mathcal{T}^2 = 12 M_{Pl}^4 \left(\frac{z}{e^{zt}-2}\right)^4 \, ,
\end{equation}
and taking into account Equations~\eqref{result 1} and \eqref{result 2} provides the following expression for $\dot{\mathcal{T}}^2$
\begin{equation}
    \dot{\mathcal{T}}^2 = -48\pi^4 g_\ast^{2}T_D^8 \left[ \frac{3(\frac{3}{64})^{1/4}\eta\sqrt{10}M_{Pl}^2 + 2\pi g^{1/2}_\ast T^2_D}{90^{5/2}M_{Pl}}\right]\,.
\end{equation}

Combining these two previous results with $f(\mathcal{T}^2)=\eta \left(\mathcal{T}^2\right)^{\frac{1}{4}}$, Equations~\eqref{eq:general term T2} and  \eqref{eq:BL to B} results in the asymmetry
\begin{equation}\label{eq:nb14 fT2}
    \frac{n_b}{s}\simeq
-\frac{\hat g\,c_{\mathcal{T}^2}k_s}{2\pi\,12^{3/4}}\;
\frac{\eta\,T_D}{M_\ast^2\,M_{Pl}\,\sqrt{g_\ast}}\;
\left[
3\left(\frac{3}{64}\right)^{1/4}\eta\sqrt{10}\,M_{Pl}^2
\;+\;
2\pi\sqrt{g_\ast}\,T_D^2
\right].
\end{equation}

For the decoupling temperature, using Equations~\eqref{eq:GammaW} and  \eqref{Friedmann 1/4} one obtains
\begin{equation}\label{eq:temp_n_1_4}
    T_D \;=\; \frac{4\pi^{4}\,\sqrt{g_\ast}\,v_H^{4}}{3\sqrt{90}\;\overline{m}_\nu^{\,2}\;M_{Pl}}\,,
\end{equation}
that is the same decoupling temperature one obtains for GR yielding $T_D \simeq 8.50\times 10^{13}$ GeV. This value is well below the Weinberg-operator EFT scale in Equation~\eqref{eq:LambdaW_EFT},
\begin{equation}
    \frac{T_D}{\Lambda_W^{\rm eff}}
    \simeq 7.1\times10^{-2},
\end{equation}
so the use of the Weinberg operator is safely within its EFT regime for the $n=1/4$ model.

\begin{figure}[H]
    \includegraphics[width=0.99\textwidth]{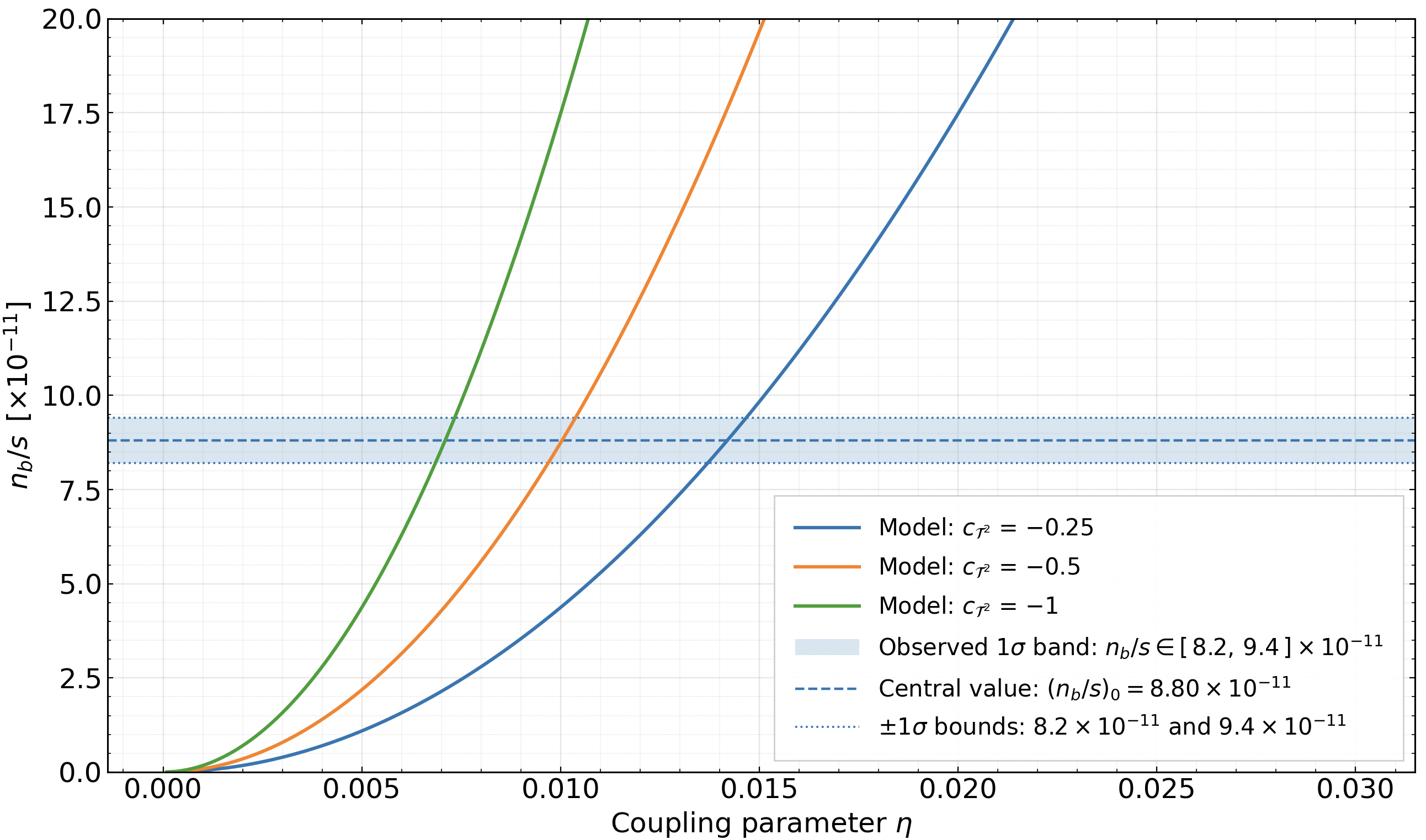}
    \caption {{Baryon}-to-entropy ratio, $\frac{n_b}{s}$, as~a function of the coupling parameter $\eta$ for three benchmark values of the coefficient $c_{\mathcal{T}^2}=\{-0.25,\,-0.5,\,-1\}$ (solid curves). The~shaded horizontal band indicates the observational $1\sigma$ range, with~the dashed line marking the central value. For~each benchmark, the~corresponding $\eta$ interval yielding agreement with observations is $[1.369799\times10^{-2},\,1.466601\times10^{-2}]$ for $c_{\mathcal{T}^2}=-0.25$, $[9.685988\times10^{-3},\,1.037040\times10^{-2}]$ for $c_{\mathcal{T}^2}=-0.5$, and~$[6.849062\times10^{-3},\,7.332996\times10^{-3}]$ for $c_{\mathcal{T}^2}=-1$. The~central-value solutions occur at $\eta=1.419028\times10^{-2}$, $1.003404\times10^{-2}$, and~$7.095135\times10^{-3}$, respectively, and~correspond to a common decoupling temperature $T_D\simeq 8.509467\times10^{13}\,\mathrm{GeV}$.}\label{Figure1}
\end{figure}

Fixing $M_\ast=M_{Pl}$ we computed numerically the evolution of the asymmetry~\eqref{eq:fRT2 term} giving the plot present in Figure~\ref{Figure1}.

\textls[-20]{The entropy interpretation of this result is non-trivial. For~$n=1/4$,
Equation~\eqref{eq:Ccon_rad_general} gives}
\begin{equation}\label{eq:Ccon_n14}
    C_{\rm con}^{(1/4)}
    =
    1+\frac{X}{2},
    \qquad
    X=
    \eta
    \left(\frac{4}{3}\right)^{-3/4}
    \left(\frac{\rho}{M_{Pl}^4}\right)^{-1/2}.
\end{equation}

{Since} $\rho\propto T^4$, one has
\begin{equation}
    X\propto T^{-2}.
\end{equation}

{Therefore}, although~the Friedmann equation in Equation~\eqref{Friedmann 1/4} takes
the standard GR form, the~radiation continuity equation does not naturally
return to the standard adiabatic limit as the Universe cools. In~fact, the~departure from adiabaticity grows at lower~temperature.

For the central-value solutions shown in Figure~\ref{Figure1}, one finds
\begin{equation}
    C_{\rm con}(T_D)
    \simeq
    \left\{
    7.90\times10^5,\,
    5.59\times10^5,\,
    3.95\times10^5
    \right\},
\end{equation}
for $c_{\mathcal T^2}=-0.25,-0.5,-1$, respectively. These values are far from the adiabatic limit $C_{\rm con}=1$. Thus, the~$n=1/4$ result in Figure~\ref{Figure1} should be interpreted as a freeze-out asymmetry rather than a final prediction for $Y_B$. If~the $n=1/4$ effective model is assumed to remain valid after baryogenesis, the~deviation from standard radiation evolution grows toward BBN. Consequently, this branch is disfavoured as a self-contained early-Universe model unless an additional transition, screening mechanism, or~UV/IR completion restores $C_{\rm con}\to1$ before~nucleosynthesis.

\subsubsection{{Model $n=1/2$}}\label{fT2 n 1/2}

For the case $n=\frac{1}{2}$, the~modified cosmological equations read
\begin{equation}\label{eq:Friedmann n=1/2}
    H^2= \left( \frac{3+\eta \sqrt{3}}{9}\right)\frac{\rho}{M_{Pl}^{2}} \, ,
\end{equation}
\begin{equation}\label{eq:Acceleration n=1/2}
   \dot{H} + H^2= -\left(\frac{3+\eta 2\sqrt{3}}{9}\right)\frac{\rho}{M_{Pl}^{2}} \,.
\end{equation}
and the scalar $\mathcal{T}^2$ and its time derivative are given by~\cite{Pereira:2025flo}
\begin{equation}\label{eq:T2 n12}
    {\mathcal{T}}^2 = \frac{4}{3}\left(\rho_{0}\right)^2 t^{-4} \, ,
 \end{equation}
\begin{equation}\label{eq:dot T2}
    \dot{\mathcal{T}}^2 = - \frac{16}{3}\left(\rho_{0}\right)^2 t^{-5} \, ,
 \end{equation}
with
\begin{equation}\label{eq:rho_0 R n=1-2}
    \rho_0(\eta) = \frac{3\sqrt{3}\eta + 9}{\left(3\eta + 2\sqrt{3}\right)^2} M_{\text{Pl}}^2 \, .
\end{equation}  

Furthermore, the~time-temperature relation for this model is given by~\cite{Pereira:2025flo}
\begin{equation}\label{eq:t_D n=1/2}
    t_D = \left(\frac{30\rho_0}{\pi^2 g_\ast}\right)^{1/2} \ T^{-2}_D \, ,
\end{equation}
that in combination with $f\left(\mathcal T^2\right) = \eta M_{Pl}^{-2} \left(\mathcal T^2\right)^{1/2}$ and Equations~\eqref{eq:T2 n12}--\eqref{eq:dot T2} gives the asymmetry
\begin{equation}
    \frac{n_b}{s}\;\simeq\;
-k_s\,\frac{\pi}{6\sqrt{10}}\;
\frac{\hat g\,c_{\mathcal{T}^2}}{M_\ast^{2}}\;
\sqrt{g_\ast}\;
\eta\;
\frac{(3\eta+2\sqrt{3})}{M_{Pl}^{3}\sqrt{3\sqrt{3}\eta+9}}\;
T_D^{5}\,,
\end{equation}
where the decoupling temperature obtained by using Equations~\eqref{eq:Friedmann n=1/2} and  \eqref{eq:GammaW} is given by
\begin{equation}
    T_D = \frac{4\pi^3v_H^4\sqrt{(3+\eta\sqrt{3})}}{9\sqrt{30}\bar{m}^2M_{Pl}}\,.
\end{equation}

Within the BBN-allowed range of the parameter $\eta$ in this model ($\eta < 0.00395156$)~\cite{Pereira:2025flo}, no asymmetry consistent with observational constraints is obtained. This outcome arises because, throughout the permitted $\eta$ interval, the~decoupling temperature is insufficiently high, while the presence of the $M_{\mathrm{Pl}}^{3}$ factor in the denominator strongly suppresses the predicted asymmetry. Consequently, the~resulting value remains well below $8.8 \times 10^{-11}$, even for moderately low choices for the cutoff scalar such as $M_\ast = 10\,T_D$, which remain within the regime of validity of the EFT description, albeit near its limit. Changing the $B\!-\!L$ source can help with this problem and lead to a result within observational~bounds.

Additionally, for~the entropy production discussion, for~$n=1/2$, the~continuity factor becomes
\begin{equation}\label{eq:Ccon_n12}
    C_{\rm con}^{(1/2)}
    =
    \frac{1+\frac{\sqrt{3}}{2}\eta}
    {1+\frac{\sqrt{3}}{3}\eta}.
\end{equation}

{This} factor is independent of temperature. Within~the BBN-allowed interval \mbox{$\eta<3.95156\times10^{-3}$}, one finds
\begin{equation}
    C_{\rm con}^{(1/2)}
    \lesssim
    1.00114.
\end{equation}

{Therefore}, the $n=1/2$ branch remains very close to the standard adiabatic radiation regime. The~entropy factor $\Delta_S$ is correspondingly close to unity, and~the usual comparison between the freeze-out asymmetry and the observed baryon asymmetry is well controlled. Hence the failure of this model to reproduce the observed value within the BBN-allowed interval is not caused by entropy dilution, but~by the smallness of the generated source~itself. 

\subsubsection{Model $n=5/8$: Adiabatic Radiation~Branch}
\label{fT2 n 5/8}

For $n=5/8$, the~radiation-era Friedmann and acceleration equations become
\begin{equation}\label{eq:Friedmann n=5/8}
    H^2
    =
    \frac{\rho}{3M_{Pl}^2}
    +
    \frac{\eta}{3M_{Pl}^{3}}
    \left(\frac{4}{3}\right)^{-3/8}
    \rho^{5/4},
\end{equation}
and
\begin{equation}\label{eq:Acceleration n=5/8}
    \dot H+H^2
    =
    -\frac{\rho}{3M_{Pl}^2}
    -
    \frac{\eta}{2M_{Pl}^{3}}
    \left(\frac{4}{3}\right)^{-3/8}
    \rho^{5/4}.
\end{equation}

{The} correction to the standard Friedmann equation is controlled by
\begin{equation}\label{eq:X_n58}
    X
    =
    \eta
    \left(\frac{4}{3}\right)^{-3/8}
    \left(\frac{\rho}{M_{Pl}^4}\right)^{1/4}.
\end{equation}

{Since} $\rho\propto T^4$, one has
\begin{equation}
    X\propto T.
\end{equation}

\textls[-15]{{Therefore}, the non-standard contribution decreases as the Universe cools. This
is qualitatively different from the $n=1/4$ branch, for~which the correction
grows at low~temperature.}

Because $C_{\rm con}^{(5/8)}=1$, the~radiation density obeys the standard
scaling
\begin{equation}
    \rho=\rho_\star\left(\frac{a_\star}{a}\right)^4.
\end{equation}

{Defining}
\begin{equation}
    x\equiv \frac{a}{a_\star},
    \qquad
    H_\star^2\equiv \frac{\rho_\star}{3M_{Pl}^2},
    \qquad
    X_\star
    \equiv
    \eta
    \left(\frac{4}{3}\right)^{-3/8}
    \left(\frac{\rho_\star}{M_{Pl}^4}\right)^{1/4},
\end{equation}
the Friedmann equation can be written as
\begin{equation}
    H^2
    =
    H_\star^2 x^{-4}
    \left(1+\frac{X_\star}{x}\right).
\end{equation}

{Hence} the scale factor is determined implicitly by
\begin{equation}\label{eq:scale_factor_n58_implicit}
    H_\star(t-t_\star)
    =
    \int_{1}^{x}
    \frac{y^{3/2}}{\sqrt{y+X_\star}}\,dy.
\end{equation}

{Equivalently}, choosing the Big Bang branch with $a=0$ at $t=0$ one gets
\begin{equation}\label{eq:scale_factor_n58_exact}
    H_\star t
    =
    \mathcal I(x;X_\star),
\end{equation}
where
\begin{equation}
\begin{aligned}
    \mathcal I(x;X_\star)
    =
    &
    \frac{x^{5/2}}{2\sqrt{x+X_\star}}
    -
    \frac{X_\star x^{3/2}}{4\sqrt{x+X_\star}}
    -
    \frac{3X_\star^2\sqrt{x}}{4\sqrt{x+X_\star}}
    \\
    &
    +
    \frac{3X_\star^2}{4}
    \sinh^{-1}
    \left(
        \sqrt{\frac{x}{X_\star}}
    \right).
\end{aligned}
\end{equation}

{In} the high-density regime $x\ll X_\star$, the~scale factor behaves as
\begin{equation}
    a(t)\propto t^{2/5}.
\end{equation}

{At} late times, $x\gg X_\star$, the~standard radiation result is recovered,
\begin{equation}
    a(t)\propto t^{1/2}.
\end{equation}

The decoupling temperature is fixed by the Weinberg interaction rate,
$\Gamma_W(T_D)=H(T_D)$. Using
\begin{equation}
    \rho=\frac{\pi^2}{30}g_\ast T^4,
\end{equation}

{Equation}~\eqref{eq:Friedmann n=5/8} gives
\begin{equation}
    H^2(T)
    =
    G_{\rm GR}T^4
    +
    \eta K_{5/8}T^5,
\end{equation}
with
\begin{equation}
    G_{\rm GR}
    =
    \frac{\pi^2 g_\ast}{90M_{Pl}^2},
    \qquad
    K_{5/8}
    =
    \frac{1}{3M_{Pl}^3}
    \left(\frac{4}{3}\right)^{-3/8}
    \left(\frac{\pi^2g_\ast}{30}\right)^{5/4}.
\end{equation}

{Writing}
\begin{equation}
    \Gamma_W(T)=D_W T^3,
    \qquad
    D_W=
    \frac{3}{4\pi^3}
    \frac{\bar m_\nu^2}{v_H^4},
\end{equation}
the freeze-out condition gives
\begin{equation}\label{eq:TD_n58}
    D_W^2T_D^2
    -
    \eta K_{5/8}T_D
    -
    G_{\rm GR}
    =
    0.
\end{equation}

{Therefore},
\begin{equation}\label{eq:TD_n58_solution}
    T_D(\eta)
    =
    \frac{
    \eta K_{5/8}
    +
    \sqrt{
    \eta^2K_{5/8}^2
    +
    4D_W^2G_{\rm GR}
    }
    }
    {2D_W^2}.
\end{equation}

{For} $\eta=0$, this reduces to the GR decoupling~temperature.

For this model,
\begin{equation}
    f(\mathcal T^2)
    =
    \eta M_{Pl}^{-3}
    \left(\mathcal T^2\right)^{5/8}\,,
\end{equation}
and during radiation domination one has
\begin{equation}
    \mathcal T^2
    =
    \frac{4}{3}\rho^2,
\end{equation}
and, therefore,
\begin{equation}
    f(\mathcal T^2)
    =
    \eta M_{Pl}^{-3}
    \left(\frac{4}{3}\right)^{5/8}
    \rho^{5/4}.
\end{equation}

{Using} $\dot\rho=-4H\rho$, one obtains
\begin{equation}\label{eq:fdot_n58}
    \partial_t f(\mathcal T^2)
    =
    -5\eta
    \left(\frac{4}{3}\right)^{5/8}
    \frac{H\rho^{5/4}}{M_{Pl}^{3}}.
\end{equation}

{Substituting} Equation~\eqref{eq:fdot_n58} into Equation~\eqref{eq:general term T2}
and using the sphaleron conversion factor in Equation~\eqref{eq:BL to B}, the~baryon-to-entropy ratio becomes
\begin{equation}\label{eq:YB_n58_general}
    \frac{n_b}{s}
    =
    -
    \frac{75\,\hat g\,k_s\,c_{\mathcal T^2}}{4\pi^2g_\ast}
    \left(\frac{4}{3}\right)^{5/8}
    \frac{\eta H_D\rho_D^{5/4}}
    {M_\ast^2 M_{Pl}^{3}T_D},
\end{equation}
where $H_D=H(T_D)$ and $\rho_D=\rho(T_D)$. Using
$H_D=\Gamma_W(T_D)=D_WT_D^3$, this can be written~as
\begin{equation}\label{eq:YB_n58_eta}
    \frac{n_b}{s}
    =
    -
    \frac{75\,\hat g\,k_s\,c_{\mathcal T^2}}{4\pi^2g_\ast}
    \left(\frac{4}{3}\right)^{5/8}
    \left(\frac{\pi^2g_\ast}{30}\right)^{5/4}
    \frac{\eta D_W T_D^7}
    {M_\ast^2M_{Pl}^{3}}.
\end{equation}

\begin{figure}[H]
\includegraphics[width=0.98\textwidth]{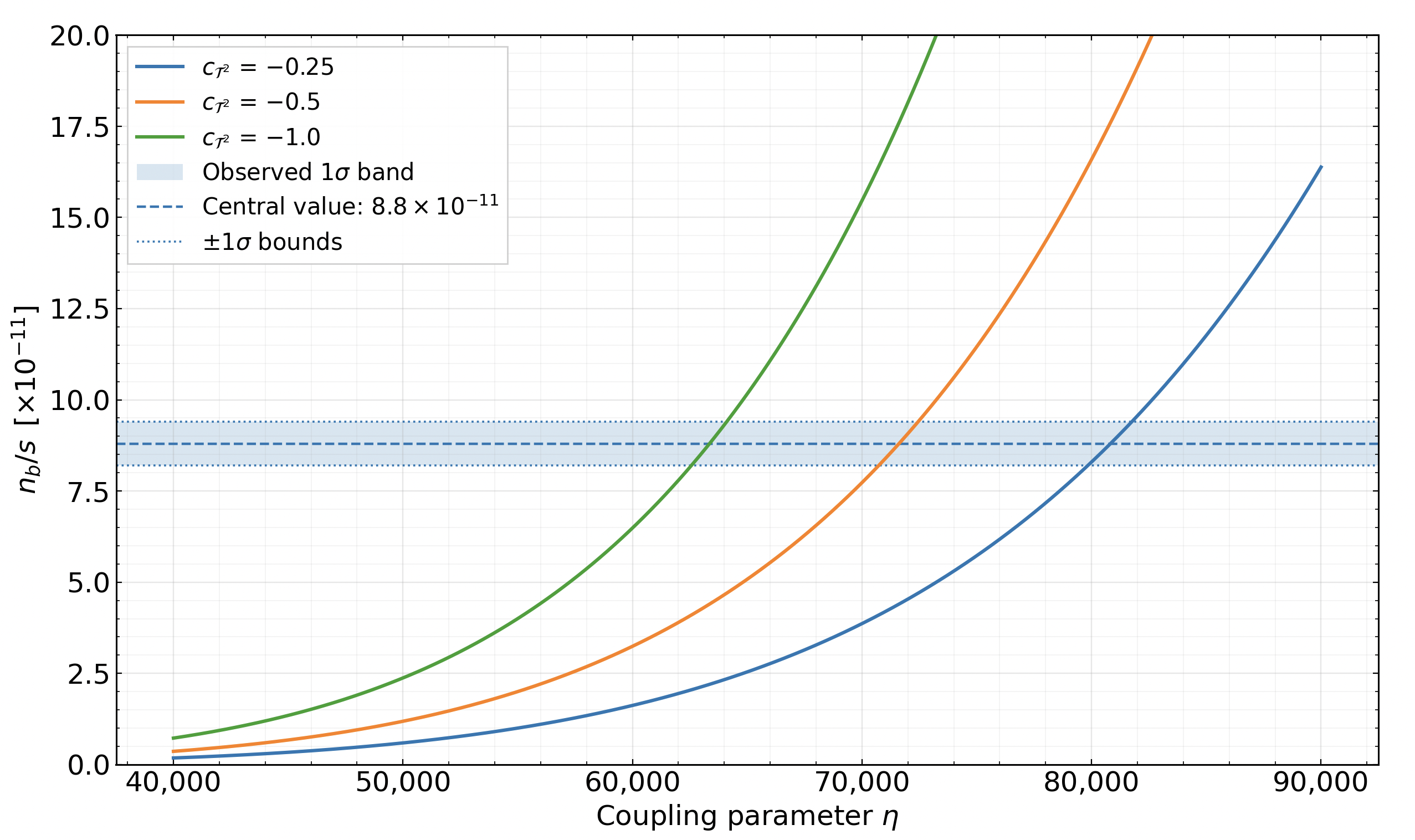}
    \caption{{Baryon}-to-entropy ratio for the entropy-safe $n=5/8$ branch as a
    function of $\eta$, using $M_\ast=10T_D$. The~shaded band denotes the
    observed $1\sigma$ interval,
    $n_b/s\in[8.2,9.4]\times10^{-11}$. Since
    $C_{\rm con}^{(5/8)}=1$, the~entropy factor is
    $\Delta_S=1$, and~the freeze-out asymmetry is equal to the final
    asymmetry. Positive baryon asymmetry is obtained for $\eta>0$ and
    $c_{\mathcal T^2}<0$.}
    \label{Figure58}
\end{figure}

{For} $\eta>0$, the~observed positive baryon asymmetry is obtained by choosing \mbox{$c_{\mathcal T^2}<0$}. In~Figure~\ref{Figure58} we show the resulting baryon-to-entropy ratio as a function of $\eta$, taking \mbox{$M_\ast=10T_D$}. This choice keeps the derivative-interaction EFT hierarchy fixed over the scan. We obtained numerically $c_{\mathcal T^2}=-1, \eta=6.7285\times10^4, T_D=4.2022\times10^{14}\,\mathrm{GeV}, M_\ast=4.2022\times10^{15}\,\mathrm{GeV}.$ For this point, $\frac{M_\ast}{T_D}=10, \frac{T_D}{\Lambda_W^{\rm eff}}\simeq0.354, \Delta_S^{(5/8)}=1,$
and $\frac{n_b}{s} \simeq 8.8\times10^{-11}$. Thus, the $n=5/8$ branch can generate the observed baryon asymmetry with healthy entropy evolution and a controlled derivative-interaction EFT~hierarchy.

The $n=5/8$ model is also BBN safe. The~change this model does to the Hubble rate is controlled by
\begin{equation}
    \frac{H^2}{H_{\rm GR}^2}=1+X,
    \qquad
    X=
    \eta
    \left(\frac{4}{3}\right)^{-3/8}
    \left(\frac{\rho}{M_{Pl}^4}\right)^{1/4}.
\end{equation}

{Since} $X\propto T$, the~correction is strongly suppressed at MeV
temperatures. Taking $T_{\rm BBN}\sim1\,\mathrm{MeV}$ and
$g_\ast\simeq10.75$, the~benchmark value
$\eta\simeq6.7\times10^4$ gives
\begin{equation}
    X_{\rm BBN}\sim10^{-17}.
\end{equation}

{Therefore},
\begin{equation}
    \frac{\Delta H}{H_{\rm GR}}
    =
    \sqrt{1+X_{\rm BBN}}-1
    \simeq \frac{X_{\rm BBN}}{2}
    \sim10^{-17},
\end{equation}
which is completely negligible for BBN. Hence the $n=5/8$ benchmark is not
constrained by the BBN expansion-rate~bound.

\unskip

\subsubsection{{Model $n=1$}}\label{fT2 n 1}

For $n=1$, the~cosmological dynamics are governed by the modified Friedmann and acceleration equations~\cite{Pereira:2025flo}
\begin{equation}\label{eq:Friedmann n=1}
	H^2 = \frac{1}{3M_{Pl}^2}\rho + \frac{2}{3}\frac{\eta}{M_{Pl}^{6}} \rho^2 \, ,  
\end{equation}
\begin{equation}\label{eq:Acc n=1}
	\dot{H} + H^2 = -\frac{1}{3M_{Pl}^2}\rho - \frac{2}{3}\frac{\eta}{M_{Pl}^{6}} \rho^2 \, ,  
\end{equation}
which lead to the expansion rate~\cite{Pereira:2025flo}
\begin{equation}\label{eq:H n=1}
	H(t) = \frac{1}{2}\,t^{-1} \, .
\end{equation}

For the scalar $\mathcal{T}^2$, one finds~\cite{Pereira:2025flo}
\begin{equation}
	\mathcal{T}^2 = \frac{\eta^{-1}M_{Pl}^6}{2}\, t^{-2} \,,
\end{equation}
\begin{equation}
	\dot{\mathcal{T}}^2 = -\eta^{-1}M_{Pl}^6\, t^{-3} \,,
\end{equation}
together with the time--temperature relation
\begin{equation}\label{decoupling time T^2 n=1}
	t_D = \sqrt{\frac{3}{8\eta}}\,30\,\pi^{-2}\,g_\ast^{-1}\,M_{Pl}^3\,T_D^{-4} \, .
\end{equation}

Combining the expressions above, the~resulting baryon asymmetry can be written as
\begin{equation}
	\frac{n_b}{s}\simeq
	-k_s\,\hat g\,c_{\mathcal{T}^2}\;
	\frac{\pi^4}{7200}\left(\frac{8\eta}{3}\right)^{3/2}
	\frac{g_*^{2}\,T_D^{11}}{M_*^2\,M_{Pl}^{9}} \, .
\end{equation}

The decoupling temperature follows from Equation~\eqref{eq:H n=1} together with Equation~\eqref{eq:GammaW}, yielding
\begin{equation}
	{T_D(\eta)=\frac{30\,\bar m_\nu^{2}\sqrt{3}\,M_{Pl}^3}{2\sqrt{8\eta}\,\pi^2 g_\ast v_H^4}\,.}
\end{equation}

The baryon asymmetry exhibits an inverse dependence on the coupling parameter $\eta$ once the decoupling condition is imposed.  Indeed the asymmetry generated at decoupling takes the schematic form
\begin{equation}
	\frac{n_b}{s}\;\propto\;\eta^{3/2}\,T_D^{11},
\end{equation}
where the factor $\eta^{3/2}$ arises from the time--temperature relation $t_D(\eta)$ (through $t_D^{-3}$), while the high power of $T_D$ originates from the $t^{-3}$ dependence of $\partial_t f$ evaluated at $t=t_D$. 
Crucially, the~decoupling temperature is not independent of $\eta$: in this model one finds
\begin{equation}
	T_D(\eta)\propto\eta^{-1/2}\,,
\end{equation}
and consequently,
\begin{equation}
	T_D^{11}\;\propto\;\left(\eta^{-1/2}\right)^{11}=\eta^{-11/2},
\end{equation}
yielding
\begin{equation}
	\frac{n_b}{s}\;\propto\;\eta^{3/2}\,\eta^{-11/2}=\eta^{-4}.
\end{equation}

\begin{figure}[H]
	\includegraphics[width=0.99\textwidth]{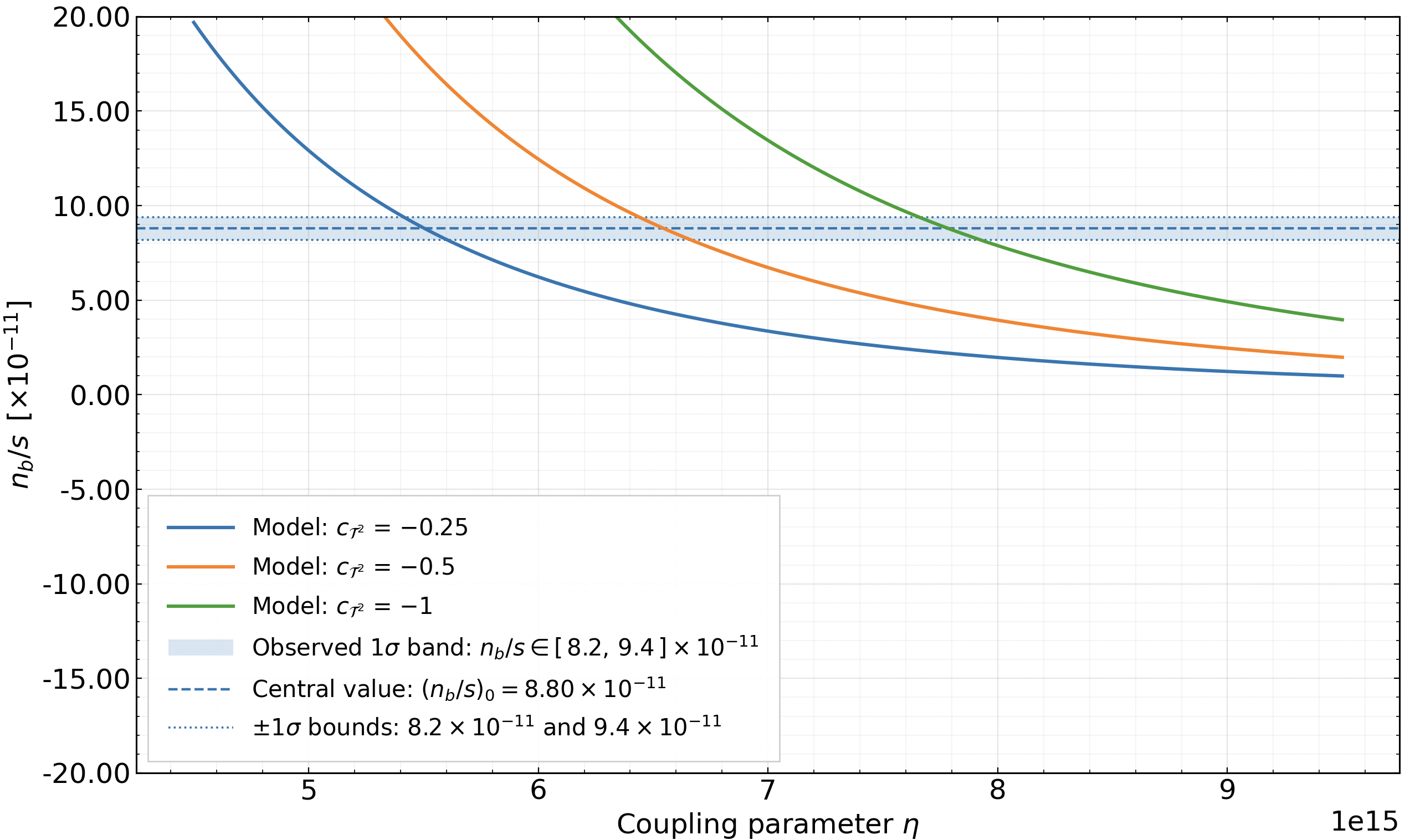}
	\caption{{Predicted} baryon-to-entropy ratio, $\frac{n_b}{s}$, as~a function of the coupling parameter $\eta$ for three representative values of the effective coefficient $c_{\mathcal{T}^2}$ (solid curves). The~light-blue shaded region denotes the observational $1\sigma$ interval, $\frac{n_b}{s} \in [8.2,\,9.4]\times 10^{-11}$, while the dashed and dotted horizontal lines indicate the central value $(\frac{n_b}{s})_0 = 8.8\times 10^{-11}$ and the corresponding $\pm 1\sigma$ bounds, respectively. The~intersections of each model curve with the central value occur at $\eta = 5.503506\times 10^{15}$ for $c_{\mathcal{T}^2}=-0.25$, $\eta = 6.544808\times 10^{15}$ for $c_{\mathcal{T}^2}=-0.5$, and~$\eta = 7.783133\times 10^{15}$ for $c_{\mathcal{T}^2}=-1$, yielding the corresponding decoupling temperatures $T_D = 1.158287\times 10^{15}\,\mathrm{GeV}$, $1.062153\times 10^{15}\,\mathrm{GeV}$, and~$9.739990\times 10^{14}\,\mathrm{GeV}$, respectively.}
	\label{Figure2}
\end{figure}

Therefore, although~$\eta$ appears explicitly as a positive power in the effective coupling, increasing $\eta$ lowers the decoupling temperature, and~the strong temperature sensitivity (here $\propto T_D^{11}$) dominates, yielding a net suppression of the final asymmetry for larger $\eta$. In~Figure~\ref{Figure2} we show $\frac{n_b}{s}$ vs $\eta$. The~figure confirms the generic trend that increasing $\eta$ suppresses the baryon-to-entropy ratio, yielding a smaller asymmetry at larger couplings. For~the representative choices of $c_{\mathcal{T}^2}$ considered, the~values of $\eta$ that reproduce the observed central value correspond to decoupling temperatures of order $T_D\sim 10^{15}\,\mathrm{GeV}$. Importantly, these parameter points are comfortably consistent with the BBN requirement, which in this model imposes the upper bound $\eta<8.87644\times 10^{82}$~\cite{Pereira:2025flo}.

\textls[-15]{The entropy correction is particularly important for this model. For~$n=1$,
\mbox{Equation~\eqref{eq:Ccon_rad_general}} gives}
\begin{equation}\label{eq:Ccon_n1}
	C_{\rm con}^{(1)}
	=
	\frac{1+2X}{1+4X},
	\qquad
	X=\eta\frac{\rho}{M_{Pl}^4}.
\end{equation}

{Unlike} the $n=1/4$ case, here $X\propto T^4$. Thus, the non-standard
contribution decreases rapidly as the Universe cools, and~the standard
adiabatic radiation regime is naturally recovered at low~temperature.

Using $X\propto T^4$ in Equation~\eqref{eq:DeltaS_general_T}, the~entropy factor
can be integrated analytically:
\begin{equation}\label{eq:DeltaS_n1}
	\Delta_S^{(1)}
	=
	\left(
	\frac{1+2X_D}{1+2X_f}
	\right)^{3/4},
\end{equation}
where $X_D=X(T_D)$ and $X_f=X(T_f)$. Once the system has evolved into the
regime $X_f\ll1$, this becomes
\begin{equation}\label{eq:DeltaS_n1_approx}
	\Delta_S^{(1)}
	\simeq
	\left(1+2X_D\right)^{3/4}.
\end{equation}

{Therefore}, the final baryon asymmetry is
\begin{equation}\label{eq:YB_final_n1}
	Y_B^{\rm final}
	=
	\frac{Y_B(T_D)}
	{\left(1+2X_D\right)^{3/4}}.
\end{equation}

For the central-value freeze-out solutions shown in Figure~\ref{Figure2}, one
finds
\begin{equation}
	C_{\rm con}(T_D)
	\simeq
	\left\{
	0.500013,\,
	0.500015,\,
	0.500018
	\right\},
\end{equation}
and
\begin{equation}
	\Delta_S
	\simeq
	\left\{
	1.67\times10^3,\,
	1.47\times10^3,\,
	1.29\times10^3
	\right\},
\end{equation}
for $c_{\mathcal T^2}=-0.25,-0.5,-1$, respectively. Hence the benchmark
points that reproduce the observed asymmetry at freeze-out would be diluted by
roughly three orders of magnitude before the standard adiabatic regime is
reached.

This dilution does not rule out the $n=1$ model, because~the asymmetry scales as $Y_B\propto M_\ast^{-2}$. Keeping the same values of $\eta$ and $c_{\mathcal T^2}$, the~observed final asymmetry can be recovered by lowering $M_\ast$ while maintaining the EFT condition $M_\ast>T_D$.  A~conservative entropy-corrected benchmark is
{\small\begin{equation}\label{eq:n1_entropy_corrected_benchmark}
	c_{\mathcal T^2}=-1,
	\qquad
	\eta = 4.37\times10^{16},
	\qquad
	T_D = 4.11\times10^{14}\,\mathrm{GeV},
	\qquad
	M_\ast = 4.11\times10^{15}\,\mathrm{GeV}.
\end{equation}}
that yields
\begin{equation}
	\frac{M_\ast}{T_D}=10,
	\qquad
	X_D\simeq1.25\times10^3,
	\qquad
	C_{\rm con}(T_D)\simeq0.5001,
	\qquad
	\Delta_S^{(1)}\simeq3.53\times10^2.
\end{equation}
with the final baryon asymmetry
\begin{equation}
	Y_B^{\rm final}
	=
	\frac{Y_B(T_D)}{\Delta_S^{(1)}}
	\simeq
	Y_B^{\rm obs}.
\end{equation}

Moreover, this benchmark satisfies the BBN~\cite{Pereira:2025flo} bound the perturbative Wilson-coefficient condition $|c_{\mathcal T^2}|\leq1$, and~the derivative-interaction EFT hierarchy $M_\ast/T_D=10$. Furthermore, using
\begin{equation}
	\Lambda_W^{\rm eff}
	=
	\frac{v_H^2}{\sqrt{\bar m_\nu^{\,2}}}
	\simeq1.2\times10^{15}\,\mathrm{GeV},
\end{equation}
one finds
\begin{equation}
	\frac{T_D}{\Lambda_W^{\rm eff}}
	\simeq0.34.
\end{equation}

{Thus}, the decoupling temperature remains below the effective Weinberg scale,
while the derivative baryogenesis operator is sufficiently suppressed for the
EFT expansion to remain under perturbative control. Therefore, after~including
the entropy dilution factor, the~$n=1$ branch remains viable for
$M_\ast<M_{Pl}$ while maintaining $M_\ast/T_D=10$.

\subsection{$f(R,\mathcal{T}^2)$ Interaction~Term}\label{subsec:generalC}

\subsubsection{{Model $n=1/4$}}\label{T^2 + R n=1/4}

Combining the results from Section \ref{fT2 n 1/4} and
\begin{equation}\label{R dot n=1/4}
    \dot{R} = 6\left[\frac{\eta\sqrt{3}M_{Pl} e^{zt} }{192^{1/4}}\left(\frac{z}{e^{zt}-2}\right)^2 \right] \, .
\end{equation}
obtained by using $\dot{R} = 6 (\ddot{H} + 4H \dot{H})$ and Equation~\eqref{H solution n=1/4} the expression from the $\partial_\mu f(R,T)$ for the $n=1/4$ model is given by
\begin{eqnarray}
    \frac{n_b}{s} \simeq \frac{\hat{g} k_s c_{\mathcal{T}^2} \pi \eta T_D (\sqrt{3}- 2(192)^{1/4})}{2(192)^{1/4}g_\ast^{1/2}M_{Pl} M_\ast^2} \left[ \left(3\eta \sqrt{10} \left(\frac{3}{64}\right)^{1/4}M_{Pl}^2 + 2\pi g^{1/2}_\ast T_D^2 \right) \right] \,,
\end{eqnarray}
with $T_D$ being given by Equation~\eqref{eq:temp_n_1_4}.

\begin{figure}[H]
    \includegraphics[width=0.99\textwidth]{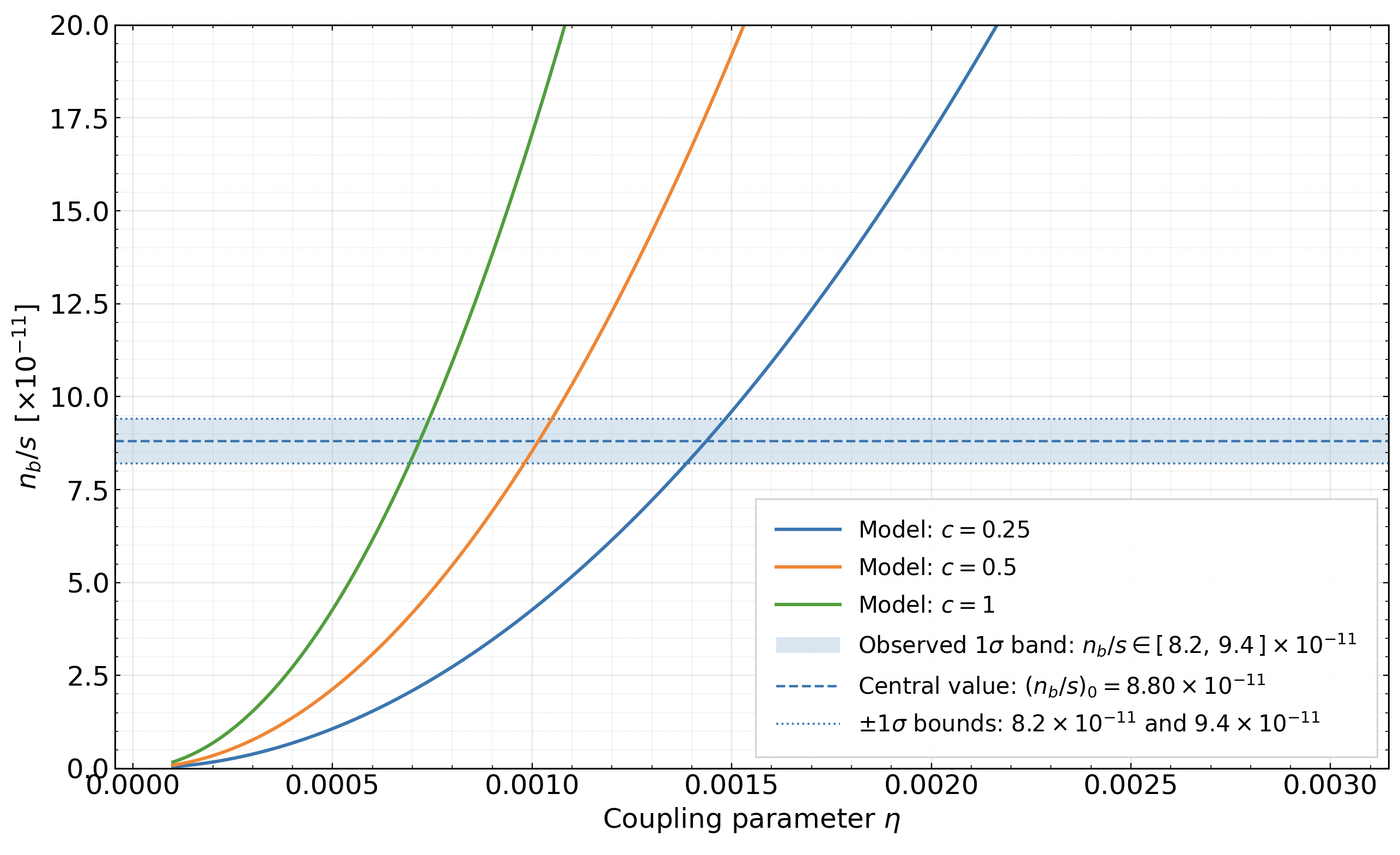}
    \caption{{Predicted} baryon-to-entropy ratio, $\frac{n_b}{s}$, as~a function of the coupling parameter $\eta$ for three representative values of the effective coefficient $c_{\mathcal{T}^2}$ (solid curves).  The~shaded horizontal band marks the observational $1\sigma$ interval. Requiring $\frac{n_b}{s}$ to fall within this band selects the parameter regions $\eta\in[1.386127\times10^{-3},\,1.484075\times10^{-3}]$ for $c=0.25$, $\eta\in[9.801399\times10^{-4},\,1.049392\times10^{-3}]$ for $c=0.5$, and~$\eta\in[6.930675\times10^{-4},\,7.420342\times10^{-4}]$ for $c=1$ The corresponding central-value solutions are located at $\eta=1.435933\times10^{-3}$, $1.015356\times10^{-3}$, and~$7.179623\times10^{-4}$, respectively.}\label{Figure4}
\end{figure}

Figure~\ref{Figure4} displays the dependence of the baryon-to-entropy ratio, $n_b/s$, on~the coupling parameter $\eta$. The~overall trend of $n_b/s(\eta)$ in the presence of the general coupling closely resembles that obtained for the pure $f(\mathcal{T}^2)$ scenario shown in Figure~\ref{Figure1}. Nevertheless, including the Ricci-scalar contribution to the asymmetry reduces the value of $\eta$ required to reproduce the observed baryon abundance. As~a representative example, for~$|c_{\mathcal{T}^2}|=0.5$ the $f(\mathcal{T}^2)$ coupling requires $\eta = 1.003404\times 10^{-2}$, whereas the combined $f(R,\mathcal{T}^2)$ coupling yields the central observational value already at $\eta = 1.015356\times 10^{-3}$, i.e.\ approximately one order of magnitude smaller. This shift suggests that the joint interaction between curvature effects and higher-order matter terms can enhance the efficiency of baryogenesis while keeping the additional gravitational-sector contribution parametrically~small.

The entropy interpretation of the $f(R,\mathcal T^2)$ interaction is governed
by the same background continuity factor as in the pure $f(\mathcal T^2)$
case. Therefore, the presence of the Ricci-scalar contribution in the
baryogenesis source does not remove the non-adiabatic behaviour of the
$n=1/4$ background. For~the central-value solutions shown in
Figure~\ref{Figure4}, one obtains
\begin{equation}
    C_{\rm con}(T_D)
    \simeq
    \left\{
    8.00\times10^4,\,
    5.65\times10^4,\,
    4.00\times10^4
    \right\},
\end{equation}
for $c_{\mathcal T^2}=0.25,0.5,1$, respectively. These values are again far
from $C_{\rm con}=1$. Hence, although~the combined $f(R,\mathcal T^2)$
interaction reduces the value of $\eta$ required to generate the freeze-out
asymmetry, it does not restore standard adiabatic radiation evolution. The \mbox{$n=1/4$} branch, therefore, remains disfavoured as a self-contained
early-Universe model unless a transition or screening mechanism restores
$C_{\rm con}\to1$ before~BBN.

\subsubsection{{Model $n=1/2$}}\label{T^2 + R n=1/2}

For $n=1/2$, using the results from Sections~\ref{fT2 n 1/2} and \ref{fT2 n 1} the asymmetry is given by
\begin{align}
\frac{n_b}{s} &\simeq
\frac{15\hat{g} k_s g_\ast^{1/2} c_{\mathcal{T}^2} \pi\eta\left(3\eta+2\sqrt{3}\right)}{M_\ast^5}
\left(
\frac{4\pi^3 v_H^4\sqrt{3+\eta\sqrt{3}}}
{9\sqrt{30}\bar{m}^2M_{Pl}}
\right)^5
\times\nonumber\\
&\left[
\frac{3\left(\eta+\sqrt{3}\right)}
{\left(30\left(3\sqrt{3}\eta+9\right)\right)^{3/2}}
+
\frac{1}
{\left(10\left(3\sqrt{3}\eta+9\right)\right)^{1/2}}
\right] .
\end{align}
  
Once more, for~the BBN-allowed range for $\eta$~\cite{Pereira:2025flo} there is no generated asymmetry in agreement with observations. The~additional contributions from the Ricci scalar to the asymmetry generation are not enough to reach a result that is in agreement with both the BBN and the observed asymmetry. The~entropy conclusion for this branch is the same as in the pure $f(\mathcal T^2)$ interaction. Since the background corresponds to $n=1/2$, the~continuity factor is given by Equation~\eqref{eq:Ccon_n12}. In~the BBN-allowed interval it remains close to unity, $C_{\rm con}\lesssim1.00114$. Therefore, entropy evolution does not change the main conclusion: the asymmetry remains below the observed value in the allowed parameter region, even after including the Ricci-scalar contribution to the baryogenesis~source.

\subsubsection{Model $n=5/8$}
\label{T^2 + R n=5/8}

We now consider the combined interaction proportional to
$\partial_\mu f(R,\mathcal T^2)J^\mu$ for the entropy-safe branch
$n=5/8$. The~background dynamics and the decoupling temperature are the same
as in Section~\ref{fT2 n 5/8}, since they are fixed by the gravitational field
equations and by the condition $\Gamma_W(T_D)=H(T_D)$. The~only difference is
the baryogenesis source, which now contains the derivative of the full
combination $R+f(\mathcal T^2)$.

For $n=5/8$, the~Ricci scalar does not vanish during radiation domination.
Using the background equations derived above, one obtains
\begin{equation}\label{eq:R_n58_RT2}
    R
    =
    -
    \eta
    \left(\frac{4}{3}\right)^{-3/8}
    \frac{\rho^{5/4}}{M_{Pl}^{3}}.
\end{equation}

{On} the other hand,
\begin{equation}
    f(\mathcal T^2)
    =
    \eta M_{Pl}^{-3}
    \left(\mathcal T^2\right)^{5/8}
    =
    \eta
    \left(\frac{4}{3}\right)^{5/8}
    \frac{\rho^{5/4}}{M_{Pl}^{3}},
\end{equation}
where we used $\mathcal T^2=4\rho^2/3$ for radiation. Hence
\begin{equation}\label{eq:R_plus_f_n58_RT2}
\begin{aligned}
    R+f(\mathcal T^2)
    &=
    \eta
    \left[
    \left(\frac{4}{3}\right)^{5/8}
    -
    \left(\frac{4}{3}\right)^{-3/8}
    \right]
    \frac{\rho^{5/4}}{M_{Pl}^{3}}
    \\
    &=
    \frac{\eta}{4}
    \left(\frac{4}{3}\right)^{5/8}
    \frac{\rho^{5/4}}{M_{Pl}^{3}}.
\end{aligned}
\end{equation}

{Thus}, for~the same values of $\eta$, $M_\ast$, and~$c_{\mathcal T^2}$, the~$f(R,\mathcal T^2)$ source is suppressed by a
factor of $1/4$ relative to the pure $f(\mathcal T^2)$ source.

Taking the time derivative of Equation~\eqref{eq:R_plus_f_n58_RT2}, one finds
\begin{equation}\label{eq:dt_R_plus_f_n58_RT2}
    \partial_t\left[R+f(\mathcal T^2)\right]
    =
    -\frac{5\eta}{4}
    \left(\frac{4}{3}\right)^{5/8}
    \frac{H\rho^{5/4}}{M_{Pl}^{3}}.
\end{equation}

{Substitution} into the baryogenesis formula gives
\begin{equation}\label{eq:YB_RT2_n58}
    \frac{n_b}{s}
    =
    -
    \frac{75\,\hat g\,k_s\,c_{\mathcal T^2}}{16\pi^2g_\ast}
    \left(\frac{4}{3}\right)^{5/8}
    \frac{\eta H_D\rho_D^{5/4}}
    {M_\ast^2M_{Pl}^{3}T_D}.
\end{equation}

{Using} $H_D=\Gamma_W(T_D)=D_WT_D^3$ and
$\rho_D=(\pi^2g_\ast/30)T_D^4$, this becomes
\begin{equation}\label{eq:YB_RT2_n58_eta}
    \frac{n_b}{s}
    =
    -
    \frac{75\,\hat g\,k_s\,c_{\mathcal T^2}}{16\pi^2g_\ast}
    \left(\frac{4}{3}\right)^{5/8}
    \left(\frac{\pi^2g_\ast}{30}\right)^{5/4}
    \frac{\eta D_W T_D^7}
    {M_\ast^2M_{Pl}^{3}}.
\end{equation}

{Therefore}, for~$\eta>0$, a~positive baryon asymmetry requires
$c_{\mathcal T^2}<0$, as~in the pure $f(\mathcal T^2)$ case.

Since the source is reduced by a factor of $1/4$, the~same asymmetry can be
obtained either by increasing $\eta$, which raises $T_D$, or~by lowering
$M_\ast$, which weakens the derivative-interaction EFT hierarchy. We found numerically that for $ c_{\mathcal T^2}=-1, \eta=8.5794\times10^4, T_D=5.2818\times10^{14}\,\mathrm{GeV}, M_\ast=5.2818\times10^{15}\,\mathrm{GeV}$ one gets $\frac{M_\ast}{T_D}=10,\frac{T_D}{\Lambda_W^{\rm eff}}\simeq0.445,\frac{n_b}{s}\simeq8.8\times10^{-11}$. For~these results the derivative-interaction EFT hierarchy is fixed at $M_\ast/T_D=10$, but~it places the decoupling temperature closer to the effective Weinberg scale than in the pure $f(\mathcal T^2)$ branch.

Thus, the combined $f(R,\mathcal T^2)$ interaction can also reproduce the observed asymmetry, although~it is less efficient than the pure $f(\mathcal T^2)$ interaction for this~branch.

\subsubsection{{Model $n=1$}}\label{T^2 + R n=1}

For $n=1$ an interesting behaviour occurs. Due to the form of $H(t)=\frac{1}{2}t^{-1}$ there is no contribution to the asymmetry from the Ricci scalar because, in~this case, $R = 0$. Hence the interaction term $f(R,\mathcal{T}^2)$ is  phenomenologically equal to $f(\mathcal{T}^2)$. The~entropy analysis is also identical to the pure $f(\mathcal T^2)$ case. Since $R=0$ for the radiation-era solution $H=1/(2t)$, the~baryogenesis source and the background entropy evolution are both governed by the $\mathcal T^2$ sector. This branch remains viable after including the dilution factor, provided the cutoff scale is lowered relative to $M_{Pl}$ while still satisfying $M_\ast>T_D$.

\subsection{Models with $n<1/4$ and $n>1$}\label{extreme n cases}

In addition to the three models previously analyzed, it is possible to explore the extreme cases where $n < 1/4$ and $n > 1$ within the context of gravitational baryogenesis. However, we did not consider these scenarios because they involve coupling constants with unnatural dimensions, which is why we refer to them as extreme cases. Nonetheless, we will provide a potential theoretical framework for both cases to be used in the context of gravitational~baryogenesis.

Starting with the modified Friedmann equations with $w=1/3$ one has
\begin{equation}
H^2=\frac{\rho }{3M_Pl^2}+\frac{\eta' \rho ^{2n}}{3} C_{Frd}\left(n,\frac{1}{3}\right) \, ,
\end{equation}
\begin{equation}
\dot{H} + H^2=-\frac{\rho }{3M_Pl^2}-\frac{\eta'
\rho ^{2n}}{3}C_{Acc}\left(n,\frac{1}{3}\right) \,,
\end{equation}
respectively.

Considering the energy scale of the very early universe ($\rho \rightarrow \infty$), two distinct approximations can be made based on whether $n < 1/4$ or $n > 1$. For~$n < 1/4$, the~$\rho$ term is expected to dominate over the $\rho^{2n}$ term because the exponent associated with $\rho^{2n}$ is less than 1. This allows the modified cosmological equations to be recast as those of GR. In~the case of $n > 1$, a~similar reasoning applies, but~here the dominant term is the quadratic term, $\rho^{2n}$, leading to modified Friedmann equations
\begin{equation}
H^2=\frac{\eta' \rho ^{2n}}{3} C_{Frd}\left(n,\frac{1}{3}\right) \, ,
\end{equation}
\begin{equation}
\dot{H}+H^2=-\frac{\eta'
\rho ^{2n}}{3}C_{Acc}\left(n,\frac{1}{3}\right) \, .
\end{equation}

It is important to emphasize that these results are valid only for sufficiently large values of the energy density. If~this condition is not met, the~parameter $\eta'$ invalidates both sets of results. Additionally, a~key point to consider is the applicability of the result derived for $n > 1$ to the specific case of $n = 1$. In~this study, we did not consider $n = 1$ as an extreme case due to the unique properties that this model exhibits. Supporting this reasoning, the~range of decoupling temperatures considered, which occur after inflation and the era of grand unification, do not satisfy the required energy density condition. We considered that for $n=1$ the approximation considered only holds when one takes into account pre-inflation or inflation as the epochs of~study.

\section{Summary and~Discussion}\label{sec:Summary}

In this work we investigated modified gravitational baryogenesis in energy-momentum squared gravity, $f(R,\mathcal T^2)$, with~$\mathcal T^2=T_{\mu\nu}T^{\mu\nu}$. The~central idea was to use the high-density matter-sector scalar $\mathcal T^2$, or~the full combination $f(R,\mathcal T^2)$, as~the time-dependent source responsible for generating a baryon asymmetry through derivative interactions with a baryonic or $B-L$ current. This extends the usual curvature-based gravitational baryogenesis mechanism by allowing the matter sector itself to contribute directly to the effective chemical~potential.

A key ingredient of the analysis was that the decoupling temperature was not introduced as a free parameter. Instead, $T_D$ was fixed by the freeze-out of the Weinberg $B-L$-violating operator through the condition $\Gamma_W(T_D)=H(T_D)$. This connects the baryogenesis calculation to the light-neutrino mass scale and to the modified expansion history of each model. The~resulting decoupling temperatures were also checked against the effective Weinberg scale in Equation~\eqref{eq:LambdaW_EFT}, through the consistency condition in Equation~\eqref{eq:Weinberg_EFT_condition}. For~the viable benchmark points, the~Weinberg operator remains within its EFT~domain.

Because the matter energy-momentum tensor is not generally conserved in $f(R,\mathcal T^2)$ gravity, the~usual adiabatic interpretation of the baryon-to-entropy ratio requires additional care. We, therefore, distinguished between the freeze-out value $Y_B(T_D)$ and the final baryon asymmetry, related by Equation~\eqref{eq:YB_final_general}. The~entropy factor $\Delta_S$, defined in Equation~\eqref{eq:DeltaS_general_T}, measures the post-freeze-out evolution of the comoving entropy. This distinction is essential: matching the observed value of $Y_B$ at freeze-out is not sufficient unless the entropy evolution is also under~control.

The entropy analysis provides a sharp discriminator between the different powers $n$. The~$n=1/4$ branch can generate a sizeable freeze-out asymmetry, but~it does not provide a satisfactory self-contained radiation-era model. Although~the Friedmann equation can take a standard-looking form, the~modified continuity equation does not return to the standard adiabatic limit. In~this case the non-adiabatic correction grows as the Universe cools, so the model would require an additional transition, screening mechanism, or~completion before BBN. Therefore, the~$n=1/4$ branch should be regarded as disfavoured as a complete early-Universe~scenario.

The $n=1/2$ branch has the opposite behaviour. Its continuity factor remains close to unity in the BBN-allowed region, so the standard entropy interpretation is well controlled. However, the~generated baryon asymmetry remains below the observed value in that same allowed region. Thus, the~$n=1/2$ model is thermodynamically healthy but not efficient enough to explain the observed baryon-to-entropy ratio in the setup considered~here.

The most important new result is the $n=5/8$ branch. This value is selected by the condition $C_{\rm con}=1$ in the radiation era, as~follows from Equation~\eqref{eq:Ccon_rad_general}. Consequently, the~radiation continuity equation takes its standard form, the~entropy factor satisfies $\Delta_S=1$, and~the freeze-out asymmetry coincides with the final asymmetry. Moreover, the~correction parameter scales as $X\propto T$, so the modified-gravity contribution decreases as the Universe cools and becomes negligible by the BBN epoch. The~model reproduces the observed asymmetry with a perturbative Wilson coefficient, a~controlled hierarchy $M_\ast>T_D$, and~a decoupling temperature below the effective Weinberg scale. We, therefore, identify the pure $f(\mathcal T^2)$, $n=5/8$ branch as the cleanest viable realization of modified gravitational baryogenesis found in this~work.

The $n=1$ branch also remains viable, but~only after including entropy dilution. In~this case the non-standard contribution decreases rapidly at low temperature, so the standard radiation regime is recovered naturally. However, the~freeze-out asymmetry is diluted by the entropy factor derived in Equation~\eqref{eq:YB_final_n1}. The~entropy-corrected benchmark in Equation~\eqref{eq:n1_entropy_corrected_benchmark} shows that the observed final asymmetry can still be obtained by lowering $M_\ast$ below $M_{Pl}$, while maintaining $M_\ast>T_D$ and $|c_{\mathcal T^2}|\leq1$. Thus, the~$n=1$ branch is viable, but~it is less direct than the $n=5/8$ branch because the entropy dilution must be included in the parameter~fit.

We also compared the pure $f(\mathcal T^2)$ interaction with the combined $f(R,\mathcal T^2)$ interaction. For~the $n=5/8$ branch, the~combined source is suppressed relative to the pure matter-sector source, as~shown in Equation~\eqref{eq:R_plus_f_n58_RT2}. Consequently, the~combined interaction can still reproduce the observed asymmetry, but~it requires either a larger value of $\eta$, which raises $T_D$, or~a smaller value of $M_\ast$, which weakens the derivative-interaction EFT hierarchy. For~this reason, the~pure $f(\mathcal T^2)$ interaction is the more efficient and cleaner baryogenesis channel for $n=5/8$.

Overall, the~results show that modified gravitational baryogenesis in $f(R,\mathcal T^2)$ gravity is viable only when the baryogenesis source, the~modified expansion rate, the~decoupling temperature, the~entropy evolution, and~the EFT hierarchy are treated together. In~particular, reproducing $Y_B^{\rm obs}$ alone is not enough. A~consistent solution must also satisfy the Weinberg-EFT condition in Equation~\eqref{eq:Weinberg_EFT_condition}, the~perturbative Wilson-coefficient condition $|c_{\mathcal T^2}|\leq1$, the~derivative-interaction hierarchy $M_\ast>T_D$, and~the requirement of a physically acceptable radiation~era.

The main conclusion is that the entropy evolution selects the viable powers of $\mathcal T^2$. The~$n=1/4$ branch is disfavoured, the~$n=1/2$ branch is entropy-safe but too weak, the~$n=1$ branch is viable after entropy dilution, and~the $n=5/8$ branch is both entropy-safe and capable of generating the observed baryon asymmetry. This makes the $n=5/8$ model the strongest candidate among the cases studied~here.

Future work should investigate whether the viability of the $n=5/8$ branch extends to a continuous interval of powers around $n=5/8$, and~whether other $B-L$-violating interactions lead to different preferred regions of parameter space. It would also be useful to embed the derivative baryogenesis operators in a more complete ultraviolet framework and to study their connection with reheating, inflationary constraints, and~possible collider or astrophysical
signatures.
\vspace{6pt}

\authorcontributions{Conceptualization, D.S.P., J.P.M. and F.S.N.L.; methodology, D.S.P., J.P.M. and F.S.N.L.; validation, D.S.P., J.P.M. and F.S.N.L.; formal analysis, D.S.P., J.P.M. and F.S.N.L.; investigation, D.S.P., J.P.M. and F.S.N.L.; writing—original draft preparation, D.S.P., J.P.M. and F.S.N.L.; writing—review and editing, D.S.P., J.P.M. and F.S.N.L.; supervision, J.P.M. and F.S.N.L.; project administration, J.P.M.; funding acquisition, J.P.M. and F.S.N.L. All authors have read and agreed to the published version of the manuscript.}

\funding{This research was funded by the Funda\c{c}\~{a}o para a Ci\^{e}ncia e a Tecnologia (FCT) from the research grants UID/04434/2025 and CEECINST/00032/2018.}



\dataavailability{The original contributions presented in this study are included in the article. Further inquiries can be directed to the corresponding author(s).}

\acknowledgments{{The} authors acknowledge funding from the Fundação para a Ciência e a Tecnologia (FCT) through the research grants UID/04434/2025. FSNL also acknowledges support from the Fundação para a Ciência e a Tecnologia (FCT) Scientific Employment Stimulus contract with reference CEECINST/00032/2018.} 

\conflictsofinterest{{The authors declare no conflicts of interest.}}


\begin{adjustwidth}{-\extralength}{0cm}

\reftitle{References}

\PublishersNote{}
\end{adjustwidth}
\end{document}